\documentclass[twocolumn,epjc3]{svjour3mod}

\usepackage{amsmath,amssymb,amsfonts,dcolumn,mathcomp,slashed,dsfont}
\usepackage{graphicx}
\usepackage{bm}
\usepackage{xcolor}
\usepackage{csquotes}
\usepackage{units}
\usepackage{flushend}
\usepackage{cite}
\usepackage[colorlinks,citecolor=blue,urlcolor=blue,linkcolor=blue]{hyperref}
\usepackage{ragged2e}
\usepackage{siunitx}
\usepackage{subcaption}
\usepackage{booktabs}
\usepackage{physics}
\usepackage{multirow}

\apptocmd\thebibliography\justifying{}{}

\newcommand{\Order}{\mathcal{O}}

\renewcommand{\dd}{{\rm d}}
\newcommand{\mpi}{M_{\pi}}
\newcommand{\mpn}{M_{\pi^0}}
\newcommand{\mpc}{M_{\pi^{\pm}}}
\newcommand{\mK}{M_{K}}
\newcommand{\mKc}{M_{K^\pm}}
\newcommand{\mKn}{M_{K^0}}

\renewcommand{\Im}{\text{Im}\,}

\newcommand{\eps}{\epsilon}

\newcommand{\GeV}{\,\text{GeV}}
\newcommand{\MeV}{\,\text{MeV}}
\newcommand{\fm}{\,\text{fm}}
\newcommand{\bsp}{\begin{sloppypar}}
\newcommand{\esp}{\end{sloppypar}}

\graphicspath{{./plots/fits/}{./plots/figures/}{./plots/tikz/}}

\newcommand{\toright}[1]{\hspace*{\fill}{\footnotesize{#1}}}

\allowdisplaybreaks

\begin{document}

\vspace{-5mm}

\title{\toright{\textnormal{PSI-PR-22-04}}\\[5mm]Kaon electromagnetic form factors in dispersion theory}

\author{
D.~Stamen\thanksref{Bonn,e1} 
\and
D.~Hariharan\thanksref{Bonn}%,e5}
\and
M.~Hoferichter\thanksref{Bern}%,e2}
\and
B.~Kubis\thanksref{Bonn}%,e3}
\and
P.~Stoffer\thanksref{Zurich,PSI}%,e4}
}

\thankstext{e1}{e-mail: stamen@hiskp.uni-bonn.de}
%\thankstext{e5}{e-mail: hariharan@hiskp.uni-bonn.de}
%\thankstext{e2}{e-mail: hoferichter@itp.unibe.ch}
%\thankstext{e3}{e-mail: kubis@hiskp.uni-bonn.de}
%\thankstext{e4}{e-mail: stoffer@physik.uzh.ch}

\institute{Helmholtz-Institut f\"ur Strahlen- und Kernphysik (Theorie) and
   Bethe Center for Theoretical Physics,
   Universit\"at Bonn, 
   53115 Bonn, Germany \label{Bonn}
   \and
  Albert Einstein Center for Fundamental Physics, Institute for Theoretical Physics, University of Bern, Sidlerstrasse 5, 3012~Bern, Switzerland \label{Bern} 
  \and
  Physik-Institut, Universit\"at Z\"urich, Winterthurerstrasse 190, 8057 Z\"urich, Switzerland
  \label{Zurich}
  \and
  Paul Scherrer Institut, 5232 Villigen PSI, Switzerland\label{PSI}
}

\date{}

\maketitle

\begin{abstract}
\bsp
The electromagnetic form factors of charged and neutral kaons are strongly constrained by their low-energy singularities, in the isovector part from two-pion intermediate states and in the isoscalar contribution in terms of $\omega$ and $\phi$ residues. The former can be predicted using the respective $\pi\pi\to\bar K K$ partial-wave amplitude and the pion electromagnetic form factor, while the latter parameters need to be determined from electromagnetic reactions involving kaons. We present a global analysis of time- and spacelike data that implements all of these constraints. The results 
enable manifold applications: kaon charge radii, elastic contributions to the kaon electromagnetic self energies and corrections to Dashen's theorem, kaon boxes in hadronic light-by-light (HLbL) scattering, and the $\phi$ region in hadronic vacuum polarization (HVP). Our main results are: $\langle r^2\rangle_\text{c}=0.359(3)\fm^2$, $\langle r^2\rangle_\text{n}=-0.060(4)\fm^2$ for the charged and neutral radii, $\eps=0.63(40)$ for the elastic contribution to the violation of Dashen's theorem, $a_\mu^{K\text{-box}}=-0.48(1)\times 10^{-11}$ for the charged kaon box in HLbL scattering, and $a_\mu^\text{HVP}[K^+K^-, \leq 1.05\GeV]=184.5(2.0)\times 10^{-11}$, $a_\mu^\text{HVP}[K_SK_L, \leq 1.05\GeV]=118.3(1.5)\times 10^{-11}$ for the HVP integrals around the $\phi$ resonance. 
The global fit to $\bar K K$ gives $\bar M_\phi=1019.479(5)\MeV$, $\bar \Gamma_\phi=4.207(8)\MeV$ for the $\phi$ resonance parameters including vacuum-polarization effects. 
\esp
\end{abstract}

%-----------------------------------------------------------------------------------------
\section{Introduction}
%-----------------------------------------------------------------------------------------

The simplest, most stringently constrained matrix element that describes the interaction of hadrons with the electromagnetic current $j_\mu$ is the pion vector form factor (VFF) 
\begin{equation}
\big\langle\pi^+(p_1)\pi^-(p_2)\big|j_\mu(0)\big|0\big\rangle=(p_1-p_2)_\mu F^V_\pi(s),
\label{eq:piVFF}
\end{equation}
where $s=(p_1+p_2)^2$. 
By far the dominant contribution to its unitarity relation arises from $\pi\pi$ intermediate states above the threshold $s_\text{th}=4\mpi^2$,
\begin{equation}
\label{FpiV}
\Im F^V_\pi(s)=F^V_\pi(s) \sin\delta(s) e^{-i\delta(s)}\theta(s-s_\text{th}),
\end{equation}
which strongly constrains the functional form of $F^V_\pi(s)$ in terms of the $P$-wave phase shift $\delta(s)$ for $\pi\pi$ scattering. Up to isospin-breaking and inelastic corrections, the unitarity relation can be solved in terms of the Omn\`es function~\cite{Omnes:1958hv}
\begin{equation}
\Omega(s)=\exp\left(\frac{s}{\pi}\int_{s_\text{th}}^\infty \dd s' \frac{\delta(s')}{s'(s'-s)}\right),
\end{equation}
up to a real polynomial $P(s)$,
\begin{equation}
F^V_\pi(s)=P(s)\Omega(s).
\end{equation}
Such dispersive constraints are ubiquitous in the literature as basis for increasingly precise representations of $F^V_\pi(s)$~\cite{DeTroconiz:2001rip,Leutwyler:2002hm,Colangelo:2003yw,deTroconiz:2004yzs,Ananthanarayan:2013zua,Ananthanarayan:2016mns,Hoferichter:2016duk,Hanhart:2016pcd,Colangelo:2018mtw,Ananthanarayan:2018nyx,Davier:2019can}, often motivated by the two-pion contribution to HVP. 

In this work, we apply the same strategy to the electromagnetic form factors of the kaon, with several key differences to the case of the pion. First, while the electromagnetic form factor of the $\pi^0$ vanishes due to $C$ parity, both charged and neutral kaons can couple to $j_\mu$, so that isoscalar and isovector components need to be considered. The isovector part possesses a unitarity relation similar to Eq.~\eqref{FpiV}~\cite{Blatnik:1978wj}, in that $\pi\pi$ intermediate states yield by far the biggest contribution, but the reaction is no longer elastic and $F_\pi^V(s)$ as well as the respective partial-wave amplitude for $\pi\pi\to\bar K K$ need to be provided as input. The unitarity relation for the isoscalar part receives dominant contributions from $3\pi$ and $\bar K K$ intermediate states, but in practice the corresponding spectral function is well approximated by the narrow $\omega$ and $\phi$ resonances, i.e., their pole parameters and residues. These key ideas are spelled out in more detail in Sec.~\ref{sec:formalism}, to establish the formalism upon which the remainder of this work will be based. 

While the isovector part can thus be predicted (and validated by data for $\tau^-\to K^- K_S\nu_\tau$~\cite{BaBar:2018qry}), the representation for the isoscalar part involves free parameters, most notably the residues of the $\omega$ and $\phi$ contributions. To determine these,
we perform fits to cross-section data for the charged and neutral timelike reactions $e^+e^-\to K^+K^-$~\cite{Achasov:2000am,CMD-2:2008fsu,BaBar:2013jqz,Kozyrev:2017agm} and $e^+e^-\to K_SK_L$~\cite{Achasov:2000am,CMD-2:2003gqi,CMD-3:2016nhy}, respectively, as well as spacelike data for charged-kaon--electron scattering~\cite{Dally:1980dj,Amendolia:1986ui}. The results of these fits are presented in Sec.~\ref{sec:fits}, including the comparison to the $\phi$ resonance parameters from Refs.~\cite{BaBar:2014uwz,Hoferichter:2019mqg,Hoid:2020xjs,Zyla:2020zbs}.

The resulting form factors can then be used to study a number of applications: 
\begin{enumerate}
\item 
The derivative at $s=0$ determines the charge radii, see Sec.~\ref{sec:charge}. For the charged kaon, the averages from Ref.~\cite{Zyla:2020zbs} are based on the spacelike data~\cite{Dally:1980dj,Amendolia:1986ui} only, such that the comparison illustrates the impact of the timelike data sets, as well as the dispersion-theoretical constraints on the isovector part. For the neutral kaon, constraints on the charge radius can be extracted from $K_L\to\pi^+\pi^-e^+e^-$~\cite{NA48:2003pwz,KTeV:2005eic} and electron scattering experiments~\cite{Foeth:1969it,Dydak:1976bv,Molzon:1978py}, allowing for another cross check. 
\item
The kaon form factors determine the elastic contribution to Compton scattering off the kaon, which, in turn, gives the bulk of the electromagnetic self energy via the Cottingham formula~\cite{Cottingham:1963zz}. Together with the analog formula for the pion, we can thus provide an estimate of the corrections to Dashen's theorem~\cite{Dashen:1969eg}---which maintains that the electromagnetic mass difference for the kaon coincides with the one for the pion in the chiral limit---at least for the (dominant) part that arises from elastic intermediate states. This estimate and the comparison to results from lattice QCD as well as extractions from $\eta\to 3\pi$ are presented in Sec.~\ref{sec:Dashen}.
\item
The spacelike form factors determine the kaon-box contributions to HLbL scattering in the anomalous magnetic moment of the muon $a_\mu=(g-2)_\mu/2$, corroborating previous estimates using vector meson dominance (VMD)~\cite{Aoyama:2020ynm} and Dyson--Schwinger (DS) equations~\cite{Eichmann:2019bqf,Miramontes:2021exi}, see Sec.~\ref{sec:HLbL}. 
\item
The timelike form factors reflect HVP in the vicinity of the $\phi$ resonance, in fact, we used precisely the same data sets that enter in the direct integration of $e^+e^-\to \text{hadrons}$ cross sections~\cite{Davier:2017zfy,Keshavarzi:2018mgv,Davier:2019can,Keshavarzi:2019abf}. Since our representation does not include excited states above the $\phi$, we cannot provide a complete account of the contribution up to a typical matching point to inclusive descriptions around $1.8\GeV$, but we can study the consistency of the various $\bar K K$ data sets around the $\phi$ region among themselves as well as with other hadronic reactions in which the $\phi$ parameters are measured. These aspects are studied in Sec.~\ref{sec:HVP}. 
\end{enumerate}
Finally, we summarize our findings and conclusions in Sec.~\ref{sec:summary}. 

%-----------------------------------------------------------------------------------------
\section{Formalism}
\label{sec:formalism}
%-----------------------------------------------------------------------------------------
While the electromagnetic form factors of charged and neutral kaons, $F_{K^{\pm,0}}(s)$, are defined in strict analogy to Eq.~\eqref{eq:piVFF}, it is more convenient for a dispersion-theoretical analysis to decompose them into isovector~($v$) and isoscalar~($s$) components according to
\begin{align}
    F_{K^{\pm}}(s) &= F_K^s(s) + F_K^v(s),\notag\\
    F_{K^{0}}(s) &= F_K^s(s) - F_K^v(s).
\end{align}
We will discuss both of these in turn in the following.

%-----------------------------------------------------------------------------------------
\subsection{Isovector part}
%-----------------------------------------------------------------------------------------

\begin{figure*}[t]
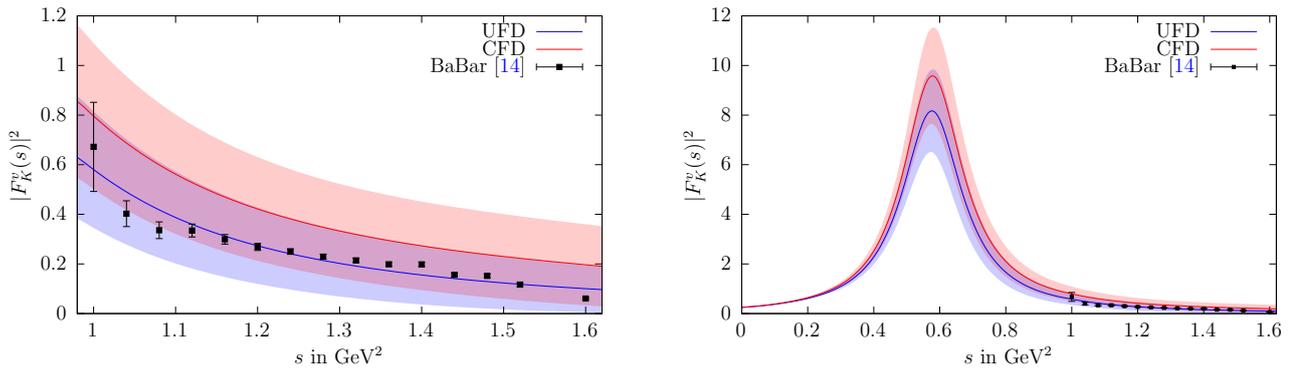

\begin{subfigure}{0.49\textwidth}
    \fontsize{12pt}{14pt} \selectfont
    \scalebox{0.65}{\input{./plots/figures/FK_compare.tex}}
    \end{subfigure}
    \hfill
    \begin{subfigure}{0.49\textwidth}
    \fontsize{12pt}{14pt} \selectfont
    \scalebox{0.65}{\input{./plots/figures/FK_compare2.tex}}
    \end{subfigure}
    \caption{The kaon isovector form factor with input from the UFD (blue) and CFD (red) parameterizations from Ref.~\cite{Pelaez:2020gnd}, compared to the BaBar data~\cite{BaBar:2018qry} for the $\tau^-\to K^-K_S\nu_\tau$ decay (left) and in the full kinematic range including the $\rho$ resonance (right).}
    \label{fig:FK_compare}
\end{figure*}

\bsp
The unitarity relation for the isovector kaon form factor reads~\cite{Blatnik:1978wj}
\begin{equation}
\Im F^v_K(s)=\frac{s}{4\sqrt{2}}\sigma^3_\pi(s)\big(g_1^{1}(s)\big)^*F_\pi^V(s). \label{eq:ImFvK}
\end{equation}
Here, $g_1^1(s)$ refers to the $\pi\pi\to\bar{K} K$ $P$-wave,
which is defined from the $\pi\pi\to \bar K K$ scattering amplitude according to~\cite{Buettiker:2003pp,Pelaez:2018qny,Pelaez:2020gnd}
\begin{align}
G^I(s,t,u)&=16\pi\sqrt{2}\sum_\ell (2\ell+1)(q_\pi q_K)^\ell P_\ell(z)g_I^\ell(s),\notag\\
q_P(s) &= \frac{\sqrt{s-4M_P^2}}{2}=\frac{\sqrt{s}}{2}\sigma_P(s),
\end{align}
where $P_\ell(z)$ are the Legendre polynomials and $z$ refers to the cosine of the scattering angle.
We specifically employ the phase of the $\pi\pi\to \bar K K$ amplitude in the Omn\`es representation for the pion VFF to render the imaginary part of Eq.~\eqref{eq:ImFvK} real by construction; this phase agrees with the pion--pion $P$-wave phase shift in the elastic region, hence this choice only affects the continuation of $\delta(s)$ above. The polynomial is fixed by a fit to the BaBar data~\cite{BaBar:2009wpw}, parameterizing the $\rho$--$\omega$ mixing via a Breit--Wigner function~\cite{Breit:1936zzb},
\begin{equation}
P(s)=1+\alpha s + \kappa \frac{s}{M_\omega^2-s-i M_\omega \Gamma_\omega},
\end{equation}
where the $\kappa$-term is dropped for the kaon form factor analysis, since the latter is performed in the isospin limit. Furthermore the polynomial is set to a constant above $\sqrt{s}=2\GeV$ to ensure convergence.
Reference~\cite{Pelaez:2020gnd} offers two alternative parameterizations of $g_1^1(s)$, an \textit{unconstrained fit to data} (UFD) of that partial wave only, as well as a variant that implements various dispersion-theoretical constraints (CFD).  
The fit parameters in the pion form factor are $\alpha_\text{UFD}=0.15(9)\GeV^{-1}$, $\alpha_\text{CFD}=0.18(8)\GeV^{-1}$, where the uncertainty is dominated by the input for $g_1^1$. In the following, the error analysis will be performed by propagating the uncertainties in the parameterization of $g_1^1$ and by linearly adding up the resulting variations, since these parameters are, in general, expected to be strongly correlated~\cite{Hoferichter:2015hva}. Since the correlations between the parameters in $g_1^1$ are not provided in Ref.~\cite{Pelaez:2020gnd}, this procedure should produce a conservative but realistic estimate of the isovector uncertainties. 
The full isovector kaon form factor can be calculated using an unsubtracted dispersion integral
\begin{equation}
F_K^v(s)=\frac{1}{\pi}\int_{s_\text{th}}^\infty \dd s' \frac{\Im F_K^v(s')}{s'-s}. \label{eq:iv-DR}
\end{equation}
Effects of higher $\rho'=\rho(1450)$ and $\rho''=\rho(1700)$ states are visible in $g_1^1$, but only affect the form factor minimally, so that the integral is dominated by the $\rho$ resonance.  This, however, only captures the effect of, e.g., the $\rho'$ partially, which is known to couple strongly to $4\pi$~\cite{Zyla:2020zbs} (see also the discussion in Ref.~\cite{Hanhart:2012wi}).  To account for intermediate states beyond $\pi\pi$ in a minimal way, we therefore add an explicit $\rho'$ resonance via a Breit--Wigner parameterization, with the coupling adjusted to fix the form factor normalization to $F^v_K(0)=1/2$. It is introduced in the form
\begin{equation}
F^{\rho'}_K(s)=\lambda_{\rho'}\frac{M_{\rho'}^2}{M_{\rho'}^2-s-i\sqrt{s}\Gamma_{\rho'}(s)},
\end{equation}
with the energy-dependent width chosen in accordance with the parameterization employed for $g_1^1$~\cite{Buettiker:2003pp,Pelaez:2018qny,Pelaez:2020gnd}\footnote{We stress that $2\pi$ and $\bar K K$ are not the dominant decay channels for $\rho'$, $\rho''$, see Ref.~\cite{Zanke:2021wiq} for more realistic spectral functions.}
\begin{equation}
	\Gamma_{\rho'}(s) = \frac{\Gamma_{\rho'} \sqrt{s}\left(2\hat{\sigma}_\pi^3(s)+\hat\sigma_K^3(s)\right)}{2M_{\rho'} \sigma_\pi^3(M^2_{\rho'})},
\end{equation}
where $\hat\sigma_P(s) = \sigma_P(s)\theta(s-4M_P^2)$.
The resulting $\rho'$ couplings are $\lambda_{\rho',\text{UFD}}=0.01(6)$ and $\lambda_{\rho',\text{CFD}}=-0.04(7)$, respectively, which demonstrates that the $\pi\pi$ intermediate states alone saturate the sum rule for the isovector charge to at least 15\% accuracy, in line with similar sum rules in Refs.~\cite{Schneider:2012ez,Hoferichter:2012pm}.
The use of an unsubtracted dispersion relation, Eq.~\eqref{eq:iv-DR}, with the addition of effective poles to satisfy the normalization constraint, guarantees a reasonable high-energy behavior of the form factor representation, $F_K^v(s) \asymp s^{-1}$ for $s\to\infty$~\cite{Leutwyler:2002hm,Chernyak:1977as,Chernyak:1980dj,Efremov:1978rn,Efremov:1979qk,Farrar:1979aw,Lepage:1979zb,Lepage:1980fj}.

\begin{figure*}[t]
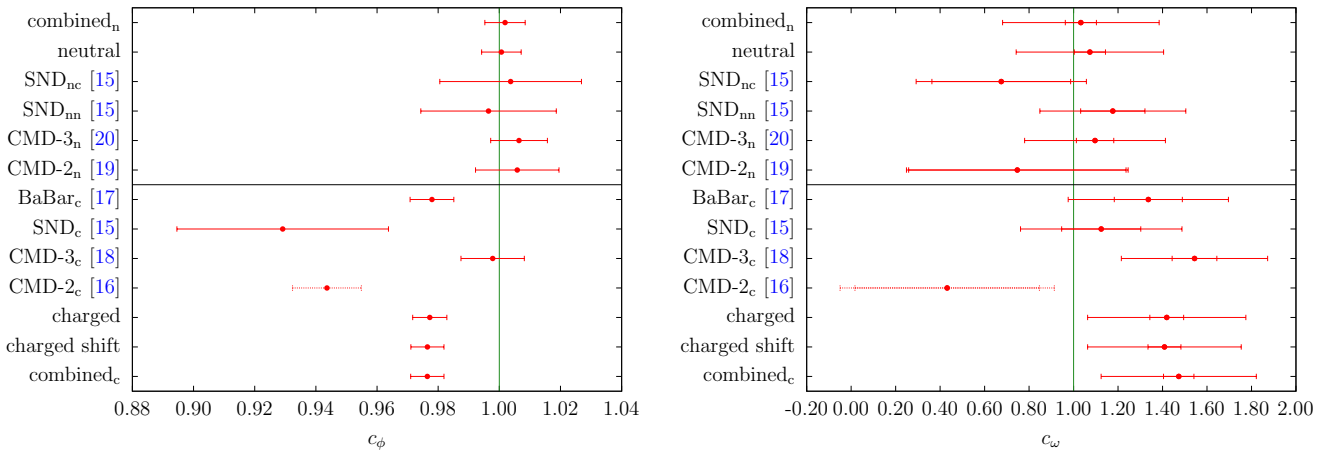

    \begin{subfigure}{0.49\textwidth}
    \fontsize{12pt}{14pt} \selectfont
    \scalebox{0.65}{\input{./plots/fits/cphi.tex}}
    \end{subfigure}
    \hfill
    \begin{subfigure}{0.49\textwidth}
    \fontsize{12pt}{14pt} \selectfont
    \scalebox{0.65}{\input{./plots/fits/comega.tex}}
    \end{subfigure}
    \caption{Results of the individual and the combined neutral and charged data sets for $c_\phi$ (left) and $c_\omega$ (right), with the neutral/charged-channel-residues in the upper/lower panel. The green line denotes the SU(3)-prediction. ``combined$_\text{n}$'' and ``combined$_\text{c}$'' refer to the scenarios in which the $\phi$ resonance parameters and all couplings are fit simultaneously, ``neutral'' and ``charged'' to the ones in which only the timelike neutral and charged data are considered, respectively, and ``charged shift'' to the variant in which a shift in the BaBar$_\text{c}$ energy calibration is allowed, see main text for details. The inner errors for $c_\omega$ refer to the fit uncertainties, the total ones are obtained by adding the systematic error from the variation of the UFD input in quadrature (a negligible effect for $c_\phi$, $M_\phi$, and $\Gamma_\phi$).}
    \label{fig:c_phi_c_omega}
\end{figure*}

Information on the isovector kaon form factor below $m_\tau$ can be obtained from $\tau^-\to K^-K_S\nu_\tau$ decays. There the spectral function $v_1(s)$ is related to the isovector form factor by
\begin{equation}
    v_1(s)=\frac{\sigma^3_K(s)}{12\pi}|F_K^v(s)|^2,
\end{equation}
up to isospin-breaking corrections, which, contrary to a determination of the $2\pi$ HVP contribution from $\tau$ decays, are not relevant at the present level of accuracy. 
A joined analysis of the $\tau^-\to K^-K_S\nu_\tau$ and $\tau^-\to\pi^-\pi^0\nu_\tau$ decays
was performed in Ref.~\cite{Gonzalez-Solis:2019iod} using resonance chiral theory in combination with dispersion relations to extract information on the $\rho'$ and $\rho''$ parameters. While the details of these higher $\rho$ excitations are not incorporated into our dispersive formalism, we expect the resulting representation to be reliable at least near threshold.   
In Fig.~\ref{fig:FK_compare} we plot the recent data from BaBar~\cite{BaBar:2018qry} against the two form factors obtained with the different inputs for $g_1^1$.
We observe that the result using the UFD input shows better agreement with the data than the CFD result. These shifts reflect the degree of consistency among the data base used in Ref.~\cite{Pelaez:2020gnd}, indicating a preference for the UFD variant.  Given that the UFD and CFD inputs agree within uncertainties, we thus opt for the more data-driven approach and adopt the UFD result in the following, to ensure better agreement with the $\tau^-\to K^-K_S\nu_\tau$ data. 
\esp

%-----------------------------------------------------------------------------------------
\subsection{Isoscalar part}
%-----------------------------------------------------------------------------------------

\begin{figure*}[t]
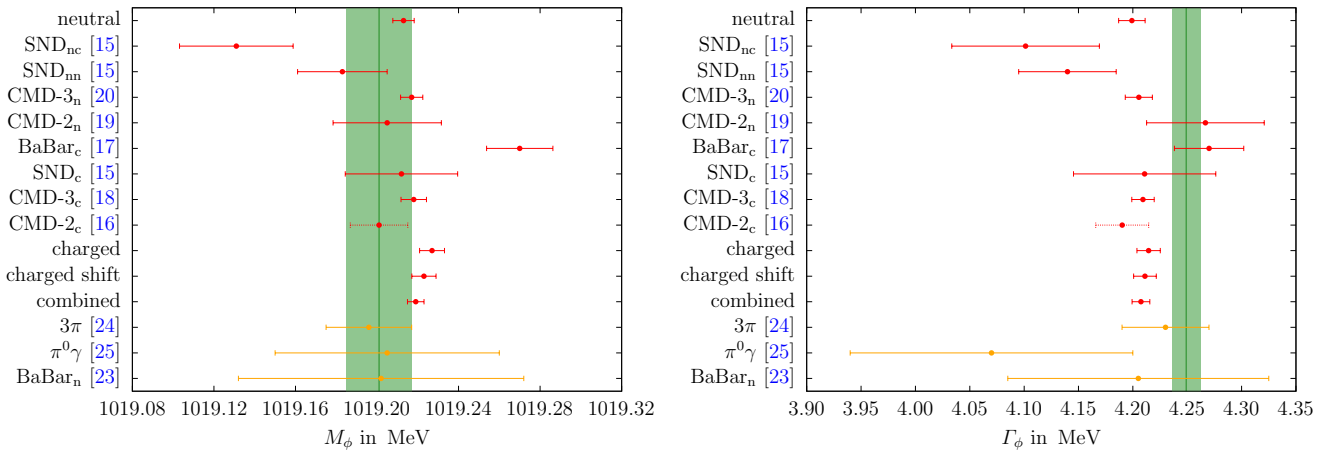

    \begin{subfigure}{0.49\textwidth}
    \fontsize{12pt}{14pt} \selectfont
    \scalebox{0.65}{\input{./plots/fits/mphi.tex}}
    \end{subfigure}
    \hfill
    \begin{subfigure}{0.49\textwidth}
    \fontsize{12pt}{14pt} \selectfont
    \scalebox{0.65}{\input{./plots/fits/Gphi.tex}}
    \end{subfigure}
    \caption{Results of the individual and the combined neutral and charged data sets for $M_\phi$ (left) and $\Gamma_\phi$ (right) in red. ``combined'' refers to the scenario in which the $\phi$ resonance parameters and all couplings are fit simultaneously, corresponding to ``combined$_\text{n}$'' and ``combined$_\text{c}$'' in Fig.~\ref{fig:c_phi_c_omega}, otherwise, the notation is as in that figure. 
    The green band corresponds to the PDG mass average~\cite{Zyla:2020zbs} shifted by the prescription from Ref.~\cite{Hoid:2020xjs} to remove the VP effect. In orange the results for the $e^+e^-\to3\pi$ and $e^+e^-\to\pi^0\gamma$ channels are given, as is the result for the neutral channel from BaBar~\cite{BaBar:2014uwz}.
    }
    \label{fig:m_phi_G_phi}
\end{figure*}

For the isoscalar part of the kaon form factors, we employ a VMD ansatz based on the lowest-lying isoscalar vector resonances $\omega(782)$ and $\phi(1020)$, as an efficient way to capture the main singularities due to $\bar K K$ and $3\pi$ intermediate states. While such a description is only strictly model-independent on the poles, the small widths of both resonances ensure that corrections beyond these dominant contributions will be appreciably suppressed. In analogy to the isovector case, we supplement this with one effective heavier pole, here chosen as the $\omega'=\omega(1420)$, to guarantee the correct form factor normalization as well as a reasonable large-$s$ behavior.  This results in
\begin{align}
\label{FK_isoscalar}
F_K^s(s)&=\frac{c_\phi}{3}\frac{M_\phi^2}{M_\phi^2-s-i\sqrt{s}\Gamma_\phi(s)}\notag\\
&\quad+\frac{c_\omega}{6}\frac{M_\omega^2}{M_\omega^2-s-iM_\omega\Gamma_\omega} \notag\\
&\quad+\left(\frac{1}{2}-\frac{c_\phi}{3}-\frac{c_\omega}{6}\right)\frac{M_{\omega'}^2}{M_{\omega'}^2-s-iM_{\omega'}\Gamma_{\omega'}},
\end{align}
in such a way that the SU(3)-symmetric limit with lowest-meson dominance corresponds to $c_\omega=c_\phi=1$. 
The energy-dependent width for the $\phi$ resonance is parameterized as~\cite{Hoferichter:2014vra}
\begin{equation}\begin{aligned}
   \Gamma_\phi(s) &= \sum_{K=K^+,K^0}\frac{\gamma_{\phi\to \bar K K}(s)}{\gamma_{\phi\to \bar K K}(M_\phi^2)}\Gamma'_{\phi\to \bar K K} \theta\big(s-4M_K^2\big) \\ &+ \frac{f_{\phi\to\pi\rho+3\pi}(s)}{f_{\phi\to\pi\rho+3\pi}(M_\phi^2)}\Gamma'_{\phi\to\pi\rho+3\pi} \theta\big(s-(M_\rho+M_\pi)^2\big),
\end{aligned}\end{equation}
where the $\Gamma'$ refer to the partial widths rescaled to compensate for all other decay channels not included explicitly, and
\begin{align}
\label{energy_dependence}
	\gamma_{\phi\to \bar K K}(s) &= \frac{\left(s-4M_K^2\right)^{3/2}}{s}, \notag\\
	f_{\phi\to\pi\rho+3\pi}(s) &= \left(\frac{\lambda\left(s,M_\rho^2,M_\pi^2\right)}{s}\right)^{3/2},
\end{align}
with $\lambda(a,b,c)=a^2+b^2+c^2-2(ab+ac+bc)$.
The widths of $\omega$ and $\omega'$  are kept constant in the timelike region for simplicity, since they only serve as smooth background terms around the $\phi$ peak and therefore cannot be further resolved in any of the data sets considered in the fit. The parameterization of the spectral functions could be further improved along the lines described in Ref.~\cite{Hoferichter:2014vra}, e.g., by a proper description of the energy dependence from the $3\pi$ channel beyond $\rho$ dominance~\eqref{energy_dependence}  or a dispersively improved variant of Eq.~\eqref{FK_isoscalar}, but in the energy regions included in the fit, i.e., the spacelike region and the timelike region around the $\phi$ peak, none of these variants would lead to any visible changes. 

%-----------------------------------------------------------------------------------------
\section{Fits to data}
\label{sec:fits}
%-----------------------------------------------------------------------------------------

\begin{figure*}[t]
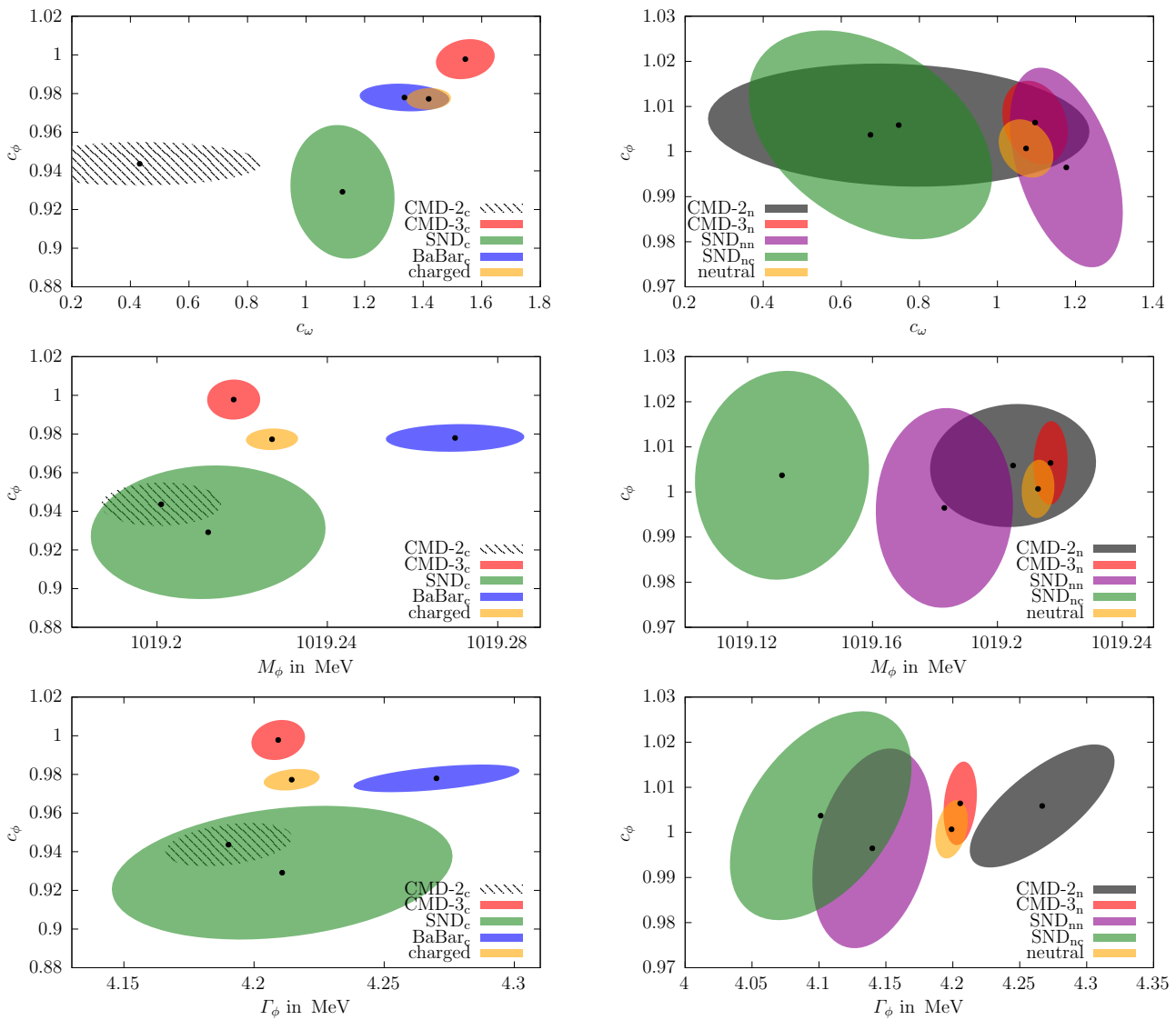

\begin{subfigure}{0.49\textwidth}
    \fontsize{12pt}{14pt} \selectfont
    \scalebox{0.65}{\input{./plots/fits/correlation_charged_3.tex}}
\end{subfigure}
\hfill
\begin{subfigure}{0.49\textwidth}
\fontsize{12pt}{14pt} \selectfont
    \scalebox{0.65}{\input{./plots/fits/correlation_neutral_3.tex}}
\end{subfigure}
\hfill
\begin{subfigure}{0.49\textwidth}
    \fontsize{12pt}{14pt} \selectfont
    \scalebox{0.65}{\input{./plots/fits/correlation_charged_1.tex}}
\end{subfigure}
\hfill
\begin{subfigure}{0.49\textwidth}
    \fontsize{12pt}{14pt} \selectfont
    \scalebox{0.65}{\input{./plots/fits/correlation_neutral_1.tex}}
\end{subfigure}
\hfill
\begin{subfigure}{0.49\textwidth}
    \fontsize{12pt}{14pt} \selectfont
    \scalebox{0.65}{\input{./plots/fits/correlation_charged_2.tex}}
\end{subfigure}
\hfill
\begin{subfigure}{0.49\textwidth}
    \fontsize{12pt}{14pt} \selectfont
    \scalebox{0.65}{\input{./plots/fits/correlation_neutral_2.tex}}
\end{subfigure}
\caption{Correlations among $c_\phi$ and $c_\omega$ (top), $M_\phi$ (middle), and $\Gamma_\phi$ (bottom), for the ``charged'' (left) and ``neutral'' (right) scenarios in comparison to the respective individual fits. The ellipses correspond to the $\Delta \chi^2=1$ contours ($39\%$ confidence level) inflated by the scale factor, in such a way that the projections reproduce the $1\sigma$ errors of the parameters. For $c_\omega$ only the fit uncertainties are shown.}
\label{fig:correlations}
\end{figure*}

\begin{figure*}[t]
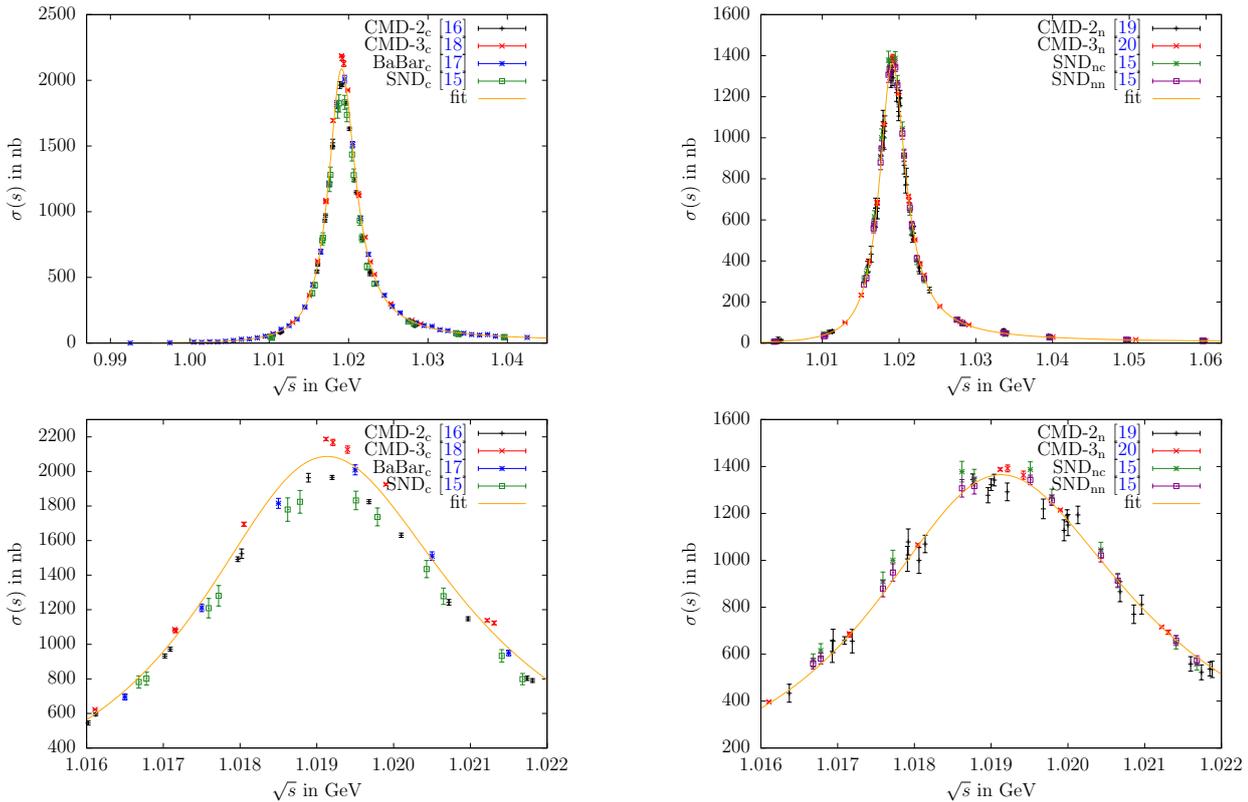

\begin{subfigure}{0.49\textwidth}
    \fontsize{12pt}{14pt} \selectfont
    \scalebox{0.6}{\input{./plots/fits/charged.tex}}
\end{subfigure}
\hfill
\begin{subfigure}{0.49\textwidth}
    \fontsize{12pt}{14pt} \selectfont
    \scalebox{0.6}{\input{./plots/fits/neutral.tex}}
\end{subfigure}
\hfill
\begin{subfigure}{0.49\textwidth}
    \fontsize{12pt}{14pt} \selectfont
    \scalebox{0.6}{\input{./plots/fits/charged_zoom.tex}}
\end{subfigure}
\hfill
\begin{subfigure}{0.49\textwidth}
    \fontsize{12pt}{14pt} \selectfont
    \scalebox{0.6}{\input{./plots/fits/neutral_zoom.tex}}
\end{subfigure}
\caption{Cross-section data as well as our combined fit to all data sets using the ``combined$_\text{c}$'' and ``combined$_\text{n}$'' residues for $e^+e^-\to K^+K^-$ (left) and $e^+e^-\to K_SK_L$ (right), respectively. Both are shown in the complete fit energy ranges (top) and focused on the $\phi$ peak region (bottom). The fits of charged and neutral channels are indistinguishable from the ``fit'' curves shown here.}
\label{fig:fits}
\end{figure*}

After establishing the formalism we now fit the resulting representations of the kaon form factors to the
available data in the time- and spacelike regions.
The timelike data are fit with the total Born cross section~\cite{Fang:2021wes}
\begin{equation}
\label{Born}
	\sigma^{(0)}(s) = \frac{\pi\alpha^2}{3s}\sigma_K^3(s) |F_K(s)|^2,
\end{equation}
multiplied by a correction factor $1+\frac{\alpha}{\pi}\eta(s)$~\cite{Hoefer:2001mx,Gluza:2002ui,Czyz:2004rj,Bystritskiy:2005ib} to account for final-state radiation (FSR). Sometimes the 
Sommerfeld--Gamow--Sakharov factor~\cite{sommerfeld1921atombau,Gamow:1928zz, Sakharov:1948plh}
\begin{equation}
\label{Gamow}
	Z(s) = \frac{\pi\alpha}{\sigma_K(s)}\frac{1+\alpha^2/\big(4\sigma_K^2(s)\big)}{1-\exp\big(-\pi\alpha/\sigma_K(s)\big)}
\end{equation}
is used instead to resum higher orders in $\alpha$, but we checked that those effects are irrelevant for the application to the $K^+K^-$ channel, and smaller than the non-Coulomb corrections contained in $\eta(s)$. 
The spacelike region is fit with the same form factor function (with all widths set to zero).
We use two sets of data obtained by $eK$ scattering for charged kaons in the spacelike region~\cite{Dally:1980dj,Amendolia:1986ui}. Data for neutral kaons in the spacelike region only exist indirectly via the scattering off atomic electrons~\cite{Foeth:1969it,Dydak:1976bv,Molzon:1978py}, leading to constraints on the neutral-kaon charge radius, and are not included in our fit.
In the timelike region, data for charged kaons are taken from CMD-2~\cite{CMD-2:2008fsu},\footnote{We show results for CMD-2~\cite{CMD-2:2008fsu} for completeness, but emphasize that these data are affected by an overestimation of the trigger efficiency for slow kaons, which leads to a systematic bias that requires a reanalysis~\cite{Novosibirsk}. Therefore, the corresponding results will be indicated by dashed lines and not included in global fits.} CMD-3~\cite{Kozyrev:2017agm}, SND~\cite{Achasov:2000am}, and BaBar~\cite{BaBar:2013jqz}. Data for neutral kaons are from CMD-2~\cite{CMD-2:2003gqi}, CMD-3~\cite{CMD-3:2016nhy}, and SND~\cite{Achasov:2000am}.\footnote{Note that BaBar~\cite{BaBar:2014uwz} do not provide cross-section data for the neutral channel in the $\phi$ region, only for the resulting $\phi$ parameters, so that their analysis cannot be included in our fit. However, the $\phi$ parameters are shown for comparison in Fig.~\ref{fig:m_phi_G_phi}.} SND has two data sets depending on the mode of detection for the $K_S$, distinguishing charged ($K_S\to\pi^+\pi^-$, referred to as SND$_\text{nc}$) and neutral mode ($K_S\to\pi^0\pi^0$, SND$_\text{nn}$). All the timelike data sets except for BaBar$_\text{c}$ and CMD-2$_\text{n}$ have vacuum-polarization (VP) effects included in the cross section. Therefore, we remove these using the routine from Ref.~\cite{Keshavarzi:2018mgv}, which outputs the running fine structure constant $\Delta\alpha(s)$. It relates the bare cross section $\sigma^{(0)}(s)$ to the dressed one $\sigma(s)$ via
\begin{equation}
    \sigma^{(0)}(s) = \sigma(s) \big| 1-\Delta\alpha(s) \big|^2.
\end{equation}
Removal of VP effects leads to a downward shift in the mass of the $\phi$ resonance by $0.260(3)\MeV$~\cite{Hoferichter:2019mqg,Hoid:2020xjs}. Crucially, this unfolding is only consistent as long as the $\phi$ masses are, which therefore needs to be monitored in the analysis below.  
In order to account for the binning in the BaBar experiment~\cite{BaBar:2013jqz}, which, in contrast to energy-scan experiments, measures the integrated signal over the bin, we take the bare cross section including FSR integrated over the bin as our fit function
\begin{equation}
\label{binaverage}
f(x_i)=\frac{1}{s_i^\text{max}-s_i^\text{min}}\int_{s_i^\text{min}}^{s_i^\text{max}} \dd s \left(1+\frac{\alpha}{\pi}\eta(s)\right)\sigma^{(0)}(s).
\end{equation}
We include the systematic uncertainties in our fits after accounting for the d'Agostini bias~\cite{DAgostini:1993arp}. In the case of strongly correlated data (most prominently for those of normalization-type origin), this bias leads to lower fit values in a chi-square minimization defined as
\begin{equation}
    \chi^2 = \sum_{i,j} \big( f(x_i)-y_i \big) V(i,j)^{-1} \big( f(x_j)-y_j \big),
\end{equation}
where $f(x)$ is the fit function, $y$ are the data points, and $V$ is the covariance matrix.
The systematic uncertainties are taken to be 100\% correlated for each experiment,\footnote{BaBar$_\text{c}$~\cite{BaBar:2013jqz} quote 9 different sources for the systematic uncertainty, to be taken as 100\% correlated individually.} and in addition fully correlated between the two detection modes SND$_\text{nc}$ and SND$_\text{nn}$~\cite{SND}.

\bsp
We follow the iterative method developed in Ref.~\cite{Ball:2009qv} to remove the bias. The modified covariance matrix is defined as
\begin{equation}
    V_{n+1}(i,j) = V^{\text{stat}}(i,j) + \frac{V^{\text{syst}}(i,j)}{y_iy_j}f_n(x_i)f_n(x_j),
\end{equation}
where the fit function and the full covariance matrix are updated in each iteration step $n$. The iteration procedure very quickly converges to the final result. Finally, the uncertainties of the fit parameters are inflated by the scale factor
\begin{equation}
\label{eq:chi2ScaleFactor}
S=\sqrt{\chi^2/\text{dof}},
\end{equation}
in case that $\chi^2/\text{dof}>1$, following the PDG prescription~\cite{Zyla:2020zbs}, to account, in a minimal way, for unknown systematic errors as indicated by the $\chi^2$.
\esp

Fits in the timelike region were constrained to energies close to the $\phi$ resonance (as shown in Fig.~\ref{fig:fits}), since we do not include higher vector resonances in the energy region above the $\phi$ with couplings adjustable to cross-section data~\cite{Beloborodov:2019fmw}. We considered several variants of the timelike input:
\begin{enumerate}
    \item\label{individual} individual fits to a single experiment (charged or neutral channel), referred to by the name of the respective experiment;
    \item\label{chargedneutral} combined fits to charged or neutral data sets, referred to as ``charged'' and ``neutral'';
    \item\label{combined} a full combination of charged and neutral data, referred to as ``combined'' (or ``combined$_\text{c}$'' and ``combined$_\text{n}$'' for the residues).
\end{enumerate}
In each scenario, the $\phi$ parameters $M_\phi$, $\Gamma_\phi$, and $c_\phi$, as well as the residue of the $\omega$ pole $c_\omega$ are allowed to float, while masses and widths of $\omega$ and $\omega'$ are kept fixed (as the fits are entirely insensitive to them). The spacelike data are included in all three variants, since, even though they are 
relatively crude in precision compared to the timelike cross-section measurements, they do help stabilize the extracted values of $c_\omega$ to some extent. We have checked that the extracted $\phi$ parameters only change within uncertainties when the SU(3) constraint $c_\omega = c_\phi$ is imposed. They are insensitive to the complete omission of the spacelike data from the fits and largely unaffected by the uncertainties from the isovector part. However, the fit value of $c_\omega$ does become sensitive to the UFD input, which we take into account by an additional systematic uncertainty, see Fig.~\ref{fig:c_phi_c_omega} and Table~\ref{tab:fit_parameters}.

In the full combination~\ref{combined} we allow $c_\omega$ and $c_\phi$ to differ in the charged and neutral channels, leading to 6 free parameters overall, since we do observe indications for isospin breaking in these parameters, see Fig.~\ref{fig:c_phi_c_omega}. Isospin violation has been studied in the past in the context of the ratio $\Gamma(\phi\to K^+K^-)/\Gamma(\phi\to K_SK_L)$~\cite{Bramon:2000qe,Flores-Baez:2008owd,Benayoun:2012etq}, finding rather small effects, with dynamical explanations tending to increase the charged-kaon coupling rather than decrease it~\cite{Bramon:2000qe}.  In principle, isospin breaking would also need to be considered when using the charged spacelike data as constraint in the fits to the neutral channel only, but in this case the data are clearly not precise enough to resolve such effects. 

\begin{table*}[t]
\renewcommand{\arraystretch}{1.3}
	\begin{tabular}{lrrrr}
		\toprule
		 & neutral & charged & charged shift & combined \\\midrule
		 $\frac{\chi^2}{\text{dof}}$ & $\frac{168}{150}=1.12$ & $\frac{139}{111}=1.25$ & $\frac{130}{110}=1.18$ & $\frac{293}{238}=1.23$\\
		 $p$-value & $15\%$ & $3.7\%$ & $9.4\%$ & $0.9\%$\\
		 $M_\phi\,[\MeV]$ & $1019.213(5)$ & $1019.227(6)$ & $1019.223(6)$ & $1019.219(4)$\\
		 $\Gamma_\phi\,[\MeV]$ & $4.199(12)$ & $4.215(11)$ & $4.211(10)$ & $4.207(8)$\\[2pt]
		 \multirow{2}{*}{$c_\phi$} & \multirow{2}{*}{$1.001(6)$} & \multirow{2}{*}{$0.977(6)$} & \multirow{2}{*}{$0.976(5)$} & $0.976(5)$\\[-2pt]
		 & & & & $1.002(7)$\\[2pt]
		 \multirow{2}{*}{$c_\omega$} & \multirow{2}{*}{$1.07(7)(42)$} & \multirow{2}{*}{$1.42(8)(35)$} & \multirow{2}{*}{$1.41(7)(34)$} & $1.47(7)(34)$\\[-2pt]
		 & & & & $1.03(7)(35)$\\[2pt]
		 $\xi\times 10^3$ & & & $-1.3(5)$ &\\
		\bottomrule
	\end{tabular}
	\centering
	\caption{Parameters for the fit variants~\ref{chargedneutral} and~\ref{combined} discussed in the main text. In the combined fit, the upper/lower values for $c_\phi$, $c_\omega$ refer to the charged/neutral channel, respectively. The second bracket for $c_\omega$ indicates the uncertainty induced by the UFD input, it is fully correlated with the corresponding variation in the isovector form factor and thus affects the neutral and charged residues in the opposite directions.}
	\label{tab:fit_parameters}
\renewcommand{\arraystretch}{1.0}
\end{table*}

\begin{figure}[t]
    \fontsize{12pt}{14pt} \selectfont
    \scalebox{0.65}{\input{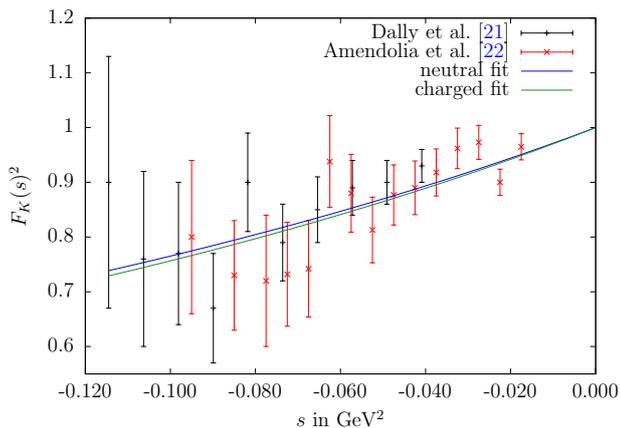}}
    \caption{Spacelike charged-kaon form factor (squared) data compared to the fits for all neutral or charged data sets combined.}
    \label{fig:space_fit}
\end{figure}

The results for the $\phi$ resonance parameters are shown in Fig.~\ref{fig:m_phi_G_phi}. In general, there is reasonable agreement among all data sets and fit variants, the exception being the BaBar$_\text{c}$ data~\cite{BaBar:2013jqz}, which favor both a larger mass and width than the remainder of the data base, as reflected also by the tension to the PDG average. This observation motivates the consideration of a variant of fit scenario~\ref{chargedneutral} in which the energy calibration in this data set is allowed to vary, referred to as ``charged shift.'' Following Ref.~\cite{Colangelo:2018mtw}, we implement such an energy shift via
\begin{equation}
\sqrt{s}\to\sqrt{s}+\xi\left(\sqrt{s}-2M_K\right)
\end{equation}
to leave the threshold invariant. On the level of the form factor this translates to a small correction~\cite{Colangelo:2018mtw}
\begin{align}
|F_K(s)|^2&\to |F_K(s)|^2\left(1+\xi A(s) + \Order(\xi^2)\right),\notag\\
A(s)&=\frac{2(s-10M_K^2)}{s+2\sqrt{s}M_K},
\end{align}
to leave the cross section~\eqref{Born} invariant, but in practice we use directly Eq.~\eqref{binaverage} in the fit. Since Ref.~\cite{BaBar:2013jqz} already provides the bare cross section, the only case in which a potential mismatch of the $\phi$ parameters when removing VP using the routine from Ref.~\cite{Keshavarzi:2018mgv} could have played a role thus remains uncritical.

The fit parameters illustrated in Figs.~\ref{fig:c_phi_c_omega} and \ref{fig:m_phi_G_phi} (and for the combined fits listed in Table~\ref{tab:fit_parameters}) display some interesting features. First, we see that there is good consistency among the neutral data sets, whose residue comes out in very good agreement with the SU(3) prediction $c_\phi=1$. In contrast, for the charged channel there is considerable spread in the fit parameters, not only in the $\phi$ mass and width, leading to the increased $\chi^2/\text{dof}$ as given in Table~\ref{tab:fit_parameters}. Moreover, the combined value of $c_\phi$ almost coincides with the BaBar$_\text{c}$ value, not the naive average with CMD-3$_\text{c}$, hinting towards an important role of correlations in the combination.  The origin of this effect is illustrated in Fig.~\ref{fig:correlations}. First, the figure reiterates the fact that consistency among the neutral data sets is much better than in the charged case, but also explains why the charged-fit value of $c_\phi$ is pulled downward compared to its naive average: $c_\phi$ is correlated with $\Gamma_\phi$ (and, to a lesser extent, $M_\phi$), and the corresponding correlations can indeed be used to reproduce the behavior of the charged fit. In this way, the tension visible in the $\phi$ resonance parameters also propagates to the extracted value of the residue. As expected, the consistency of the fit does improve slightly in the ``charged shift'' variant, with the new $\chi^2/\text{dof}$ given in Table~\ref{tab:fit_parameters}.\footnote{We note that the inclusion of the lower-precision spacelike data partly conceals the extent of the tension: without spacelike data, the $p$-value drops to $0.9\%$ in the ``charged'' fit and to $3.1\%$ in the ``charged shift'' fit, while the fit parameters remain essentially unchanged.} We observe that the residues are not affected by this shift and the mass $M_\phi$ is only shifted slightly towards the PDG value (cf.\ Figs.~\ref{fig:c_phi_c_omega} and \ref{fig:m_phi_G_phi}).
Finally, we also show the outcome of the fits in the spacelike region, see Fig.~\ref{fig:space_fit}. In particular, there is no visible difference in the neutral and charged fit,  which indicates the weak dependence on the fit parameters on this region and justifies, a posteriori, neglecting isospin breaking in the neutral fit.

An improved determination of $c_\omega$ could, in combination with the $\omega$--photon coupling constant well known from $\omega\to e^+e^-$~\cite{Klingl:1996by,Hoferichter:2017ftn}, be used to extract the coupling of the $\omega$ to kaons.  As pointed out in Ref.~\cite{Dax:2020dzg}, the latter constitutes one of the dominant uncertainties in an analysis of the reaction $\gamma K\to K\pi$, where a value deduced from a VMD analysis of kaon form factors in a wider (timelike) energy range~\cite{Beloborodov:2019fmw} was employed, corresponding to $c_\omega=1.29(15)$.  From our analysis, we conclude that the inclusion of spacelike data does not reduce the uncertainty in this coupling constant appreciably, cf.\ Table~\ref{tab:fit_parameters}; to the contrary, the uncertainties propagated from the isovector part of the form factor imply a larger uncertainty than quoted in Ref.~\cite{Beloborodov:2019fmw}. 

Adding back VP effects, our combined fit~\ref{combined} gives
\begin{align}
\label{phi_parameters_our_fit}
\bar M_\phi\big|_\text{our fit} &=1019.479(5)\MeV, \notag\\
\bar \Gamma_\phi\big|_\text{our fit} &=4.207(8)\MeV,    
\end{align}
to be compared with the PDG averages
\begin{align}
\bar M_\phi\big|_{\text{\cite{Zyla:2020zbs}}} &=1019.461(16)\MeV, \notag\\
\bar \Gamma_\phi\big|_{\text{\cite{Zyla:2020zbs}}} &=4.249(13)\MeV,    
\end{align}
see also Fig.~\ref{fig:m_phi_G_phi}. The $\phi$ mass comes out consistent within errors, but our global fit suggests a sizable reduction in uncertainty, as a result of fitting all available cross sections in a combined analysis, instead of averaging only the resonance parameters as quoted by each experiment. For the width, the resulting uncertainty comes out similarly as in Ref.~\cite{Zyla:2020zbs}, but with a central value that is lower by $2.8\sigma$. In both cases, our results from $\bar K K$ are consistent with previous extractions from the $3\pi$ and $\pi^0\gamma$ channels~\cite{Hoferichter:2019mqg,Hoid:2020xjs}. 

%-----------------------------------------------------------------------------------------
\section{Applications}
\label{sec:app}
%-----------------------------------------------------------------------------------------

%-----------------------------------------------------------------------------------------
\subsection{Charge radii of the kaon}
\label{sec:charge}
%-----------------------------------------------------------------------------------------

\begin{figure*}[t]
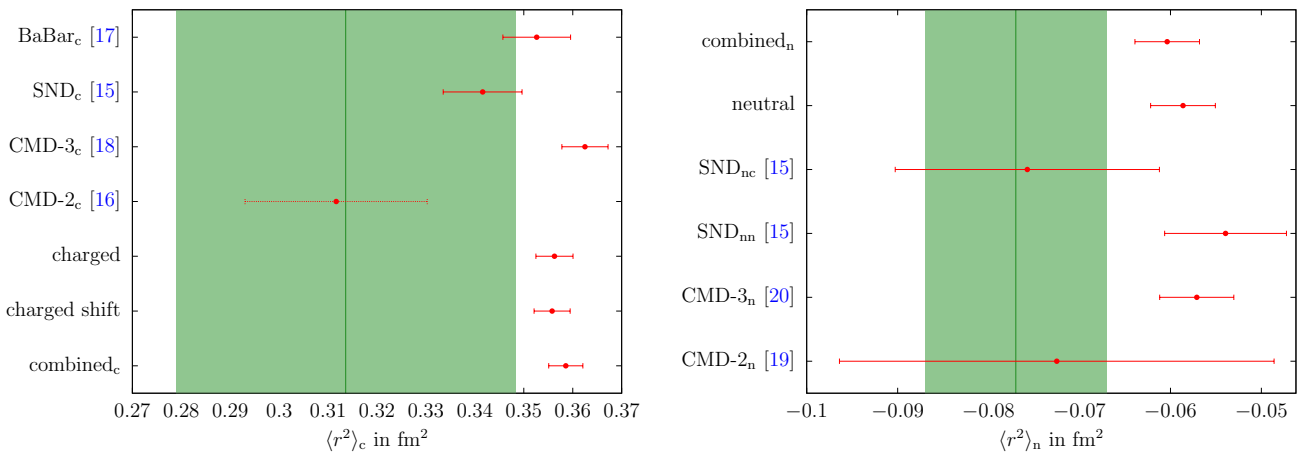
  
    \begin{subfigure}{0.49\textwidth}
        \fontsize{12pt}{14pt} \selectfont
        \scalebox{0.65}{\input{./plots/fits/chargedradius.tex}}
    \end{subfigure}
    \hfill
    \begin{subfigure}{0.49\textwidth}
        \fontsize{12pt}{14pt} \selectfont
        \scalebox{0.65}{\input{./plots/fits/neutralradius.tex}}
    \end{subfigure}  
    \caption{Results of the individual and the combined charged (left) and neutral (right) data sets of the squared mean charge radius for the charged (left) and neutral (right) kaon, for all fit variants in the same convention as in Fig.~\ref{fig:c_phi_c_omega}. The uncertainty is obtained by adding the fit uncertainty and the systematic error from the UFD input in quadrature. The green band denotes the value quoted in Ref.~\cite{Zyla:2020zbs}.}
    \label{fig:neutralandchargedradius}
\end{figure*}

As first application of the results from Sec.~\ref{sec:fits} we consider the kaon charge radii, which are defined by the derivative at $s=0$
\begin{equation}
	 \langle r^2 \rangle = 6\frac{\textrm{d}F(s)}{\textrm{d}s}\bigg|_{s=0}.
\end{equation}
The PDG averages of the neutral and charged kaon charge radii are~\cite{Zyla:2020zbs}
\begin{align}
    \langle r^2 \rangle_\text{n}\big|_{\text{\cite{Zyla:2020zbs}}} &=-0.077(10)\fm^2&&\text{\cite{Molzon:1978py,NA48:2003pwz,KTeV:2005eic}},\notag\\
    \langle r^2 \rangle_\text{c}\big|_{\text{\cite{Zyla:2020zbs}}} &=0.314(35)\fm^2&&\text{\cite{Dally:1980dj,Amendolia:1986ui}},
\end{align}
which are shown as green bands in Fig.~\ref{fig:neutralandchargedradius} and compared to our computed results, including the values from the combined fit~\ref{combined} 
\begin{align}
\label{charge_radii_main_result}
    \langle r^2 \rangle_\text{n}\big|_\text{our fit} &=-0.060(3)(2)\fm^2=-0.060(4)\fm^2,\notag\\
    \langle r^2 \rangle_\text{c}\big|_\text{our fit} &=0.359(3)(2)\fm^2=0.359(3)\fm^2.
\end{align}
Here and below, we quote the results from the combined fit as our main result, given that this implements the maximum amount of independent constraints, in particular, universality of the $\phi$ pole parameters. The first error refers to the fit uncertainties, the second one to the uncertainties due to the UFD input (fully propagated, including the indirect effect via $c_\omega$). Both sources of error can therefore be considered uncorrelated and added in quadrature.  

In the charged channel, the results are compatible within $1.3\sigma$, which is expected given that the result quoted in Ref.~\cite{Zyla:2020zbs} is calculated from the same spacelike experiments that are used in our analysis. However, the inclusion of the timelike data as well as the dispersive constraints on the isovector component allow us to improve the precision by an order of magnitude.

For the neutral kaon we also observe a sizable reduction in uncertainty, here our result lies $1.6\sigma$ higher than Ref.~\cite{Zyla:2020zbs}, whose average is dominated by the extraction~\cite{KTeV:2005eic} from $K_L\to\pi^+\pi^- e^+e^-$. The latter requires some assumptions on the other decay mechanisms not involving the kaon form factor, and it is noteworthy that the best determination from $K^0$--electron scattering~\cite{Molzon:1978py} finds a central value even higher than ours, albeit with a large uncertainty. Given the slight tension with Ref.~\cite{KTeV:2005eic}, we also studied variants of our fit in which the neutral radius from Ref.~\cite{Zyla:2020zbs} is imposed as another constraint, with minimal changes to the fit outcome.  

Finally, we can also compare to the strict VMD predictions
\begin{align}
  \langle r^2 \rangle_\text{n}\big|_\text{VMD} &=\frac{2}{M_\phi^2}+\frac{1}{M_\omega^2}-\frac{3}{M_\rho^2}\simeq -0.06\fm^2,\notag\\
    \langle r^2 \rangle_\text{c}\big|_\text{VMD} &=\frac{2}{M_\phi^2}+\frac{1}{M_\omega^2}+\frac{3}{M_\rho^2}\simeq 0.33\fm^2, 
\end{align}
which shows, a posteriori, that at least in the charge radii the deviations from VMD are small. In the same way, the related low-energy constants in chiral perturbation theory will come out close to the expectation from resonance saturation~\cite{Gasser:1984ux,Bijnens:2002hp}.

%-----------------------------------------------------------------------------------------
\subsection{Corrections to Dashen's theorem}
\label{sec:Dashen}
%-----------------------------------------------------------------------------------------

A precise determination of the electromagnetic mass difference for kaons 
\begin{equation}
    (\Delta M_K^2)_{\text{EM}} = \big(\mKc^2-\mKn^2\big)_{\text{EM}}
\end{equation}
is important for extractions of the quark mass difference $\delta=m_d-m_u$ from meson masses. The combination
\begin{equation}
    Q^2=\frac{\mK^2}{\mpi^2}\frac{\mK^2-\mpi^2}{(\mKn^2-\mKc^2)_\text{str}}\Big\{1+\Order\big(m_q^2,\delta,e^2\big)\Big\}, \label{eq:Q}
\end{equation}
which determines the major semi-axis in Leutwyler's ellipse~\cite{Leutwyler:1996qg}
\begin{equation}
    \bigg(\frac{m_u}{m_d}\bigg)^2+\frac{1}{Q^2}\bigg(\frac{m_s}{m_d}\bigg)^2=1,
\end{equation}
is particularly stable with respect to strong higher-order corrections~\cite{Gasser:1984gg}, but to make use of this relation the masses need to be corrected for their electromagnetic contributions. 
Dashen's theorem predicts~\cite{Dashen:1969eg}
\begin{equation}
\label{eq:Dashen}
(\Delta M_K^2)_{\text{EM}}=(\Delta M_\pi^2)_{\text{EM}}+\Order\big(e^2m_q\big),
\end{equation}
but corrections are large, as shown in Refs.~\cite{Bijnens:1993ae,Donoghue:1993hj,Baur:1995ig,Ananthanarayan:2004qk}. To quantify these corrections, one commonly defines the parameter
\begin{equation}
\eps=\frac{(\Delta M_K^2)_{\text{EM}}}{(\Delta M_\pi^2)_{\text{EM}}}-1.
\end{equation}
The most recent lattice-QCD averages are~\cite{Aoki:2021kgd}: 
\begin{align}
\label{lattice}
&N_f=2+1+1: &\eps&=0.79(6) &&\text{\cite{Borsanyi:2014jba,Giusti:2017dmp,MILC:2018ddw}},\notag\\
&N_f=2+1: &\eps&=0.73(17)&&\text{\cite{Fodor:2016bgu}}.
\end{align}
Phenomenologically, the electromagnetic contributions can be estimated via the Cottingham formula~\cite{Cottingham:1963zz,Harari:1966mu}, which establishes a connection between the electromagnetic self energies and the forward Compton tensor. This approach has been used extensively to separate the proton--neutron mass difference into strong and electromagnetic pieces~\cite{Gasser:1974wd,Gasser:1982ap,Walker-Loud:2012ift,Thomas:2014dxa,Erben:2014hza,Gasser:2015dwa,Gasser:2020mzy,Gasser:2020hzn}, but applies to any self-energy-type matrix element that arises from the contraction of two external currents, including meson masses~\cite{Ecker:1988te,Bardeen:1988zw,Donoghue:1993hj,Baur:1995ig,Donoghue:1996zn} or even contact-term contributions in neutrinoless double $\beta$ decay~\cite{Cirigliano:2020dmx,Cirigliano:2021qko}. In particular, the dominant contributions are typically generated by the elastic intermediate states, e.g., for the pion, in which case strong contributions to the mass difference are suppressed by $\delta^2$, one can check that the pion pole gives more than $90\%$ of the total, with small axial-vector corrections expected to make up the remainder. Since these elastic contributions are fully determined by the respective electromagnetic form factor, we can thus apply our result to obtain an estimate of the electromagnetic self energies of the kaons, and, in combination with the analog result for the pion, derive the corresponding value of $\eps$. 

\begin{figure}[t]
    \centering
    \hspace{1cm}
    \begin{subfigure}[t]{0.15\textwidth}
        \centering
        \includegraphics[width=\textwidth]{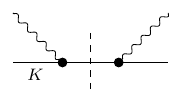}
    \end{subfigure}
    \hfill
    \begin{subfigure}[t]{0.15\textwidth}
        \centering
        \includegraphics[width=\textwidth]{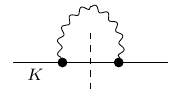}
    \end{subfigure} 
    \hspace{1cm}
	\caption{Forward scattering amplitude (left) and self-energy contraction (right). The thick black dots denote kaon electromagnetic form factor insertions, the dashed line indicates the kaon pole. Crossed diagrams are omitted.}
	\label{fig:Feynman-Dashen}
\end{figure}

To this end, we need a variant of the Cottingham formula that includes strong higher-order corrections, while the expressions in the literature are typically given in the chiral limit. Starting point is the relation
\begin{equation}
    (M_P^2)_{\text{EM}}=\frac{i e^2}{2}\int \frac{\dd^4 k}{(2\pi)^4}\frac{T^\mu_\mu}{k^2+i\epsilon},
\end{equation}
where $T^\mu_\mu$ is the contracted Compton tensor in forward direction, see Fig.~\ref{fig:Feynman-Dashen}. The elastic contribution reads
\begin{equation}
\label{Tmumu_el}
T^\mu_\mu\big|_\text{el}=\frac{2k^2(3k^2-4M_P^2)-16(k\cdot p)^2}{(k^2)^2-4(k\cdot p)^2} \big[F_P(k^2)\big]^2,   
\end{equation}
where $F_P(k^2)$ refers to the electromagnetic form factor of the meson $P$ and $p$ is its on-shell momentum. Wick-rotating $k^0$ onto the imaginary axis, the integral yields 
\begin{align}
\label{Cottingham}
(M_P^2)_\text{EM}
&=\frac{\alpha}{8\pi}\int_0^\infty \dd s \,\big[F_{P}(-s)\big]^2\notag\\
&\quad\times \Big(4W+\frac{s}{M_{P}^2}\left(W-1\right)\Big),
\end{align}
with $W=\sqrt{1+4M_P^2/s}$, and the limit $M_P\to 0$ reproduces the corresponding expressions in the literature.\footnote{For $M_P^2\ll s$ one has $4W+(W-1)s/M_{P}^2\to 6$. However, at finite $M_P$ this kernel behaves as $8M_P/\sqrt{s}$ for $s\to 0$, which explains why the kaon self-energy integral receives a larger contribution from low energies than the one for the pion.} Strictly speaking, to identify the elastic contributions~\eqref{Tmumu_el} one needs to analyze a dispersion relation and evaluate the single-particle poles, but in contrast to the nucleon case this does not lead to any subtleties and results in the scalar-QED expression multiplied with the electromagnetic form factor, see Ref.~\cite{Colangelo:2015ama}.  
As the elastic contribution to $(\mpn^2)_{\text{EM}}$ vanishes, $(\Delta M_\pi^2)_{\text{EM}} = (\mpc^2)_{\text{EM}}$ at this order, while the electromagnetic kaon mass difference may be rewritten according to
\begin{align}
\label{Cottingham-Kaon}
(\Delta M_K^2)_\text{EM}
&=\frac{\alpha}{2\pi}\int_0^\infty \dd s \,F_K^v(-s) F_K^s(-s)\notag\\
&\quad\times \Big(4W+\frac{s}{M_{K}^2}\left(W-1\right)\Big),
\end{align}
illustrating the role of the isovector kaon form factor that is theoretically particularly well constrained by the present analysis. Since we allow for residual isospin-breaking effects in our fits, which affect the isoscalar form factor, we still rely on Eq.~\eqref{Cottingham} for the numerical evaluation.

\begin{figure}[t]
    \fontsize{12pt}{14pt} \selectfont
    \scalebox{0.65}{\input{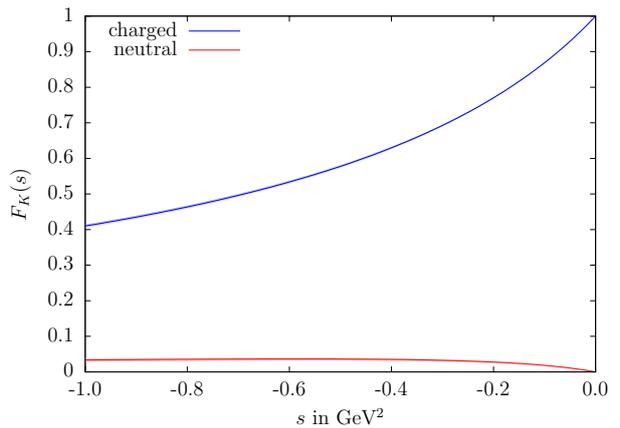}}
    \caption{The charged (blue) and neutral (red) kaon form factors in the spacelike region as determined in Sec.~\ref{sec:fits}.}
    \label{fig:F_K_full}
\end{figure}

Using our result for the kaon form factors in the spacelike region, see Fig.~\ref{fig:F_K_full}, as well as the pion form factor from Ref.~\cite{Colangelo:2018mtw}, we find
\begin{align}
(M_{K^\pm}^2)_\text{EM}&=2.12(2)(17)\times 10^{-3}\GeV^2\notag\\
&=2.12(18)\times 10^{-3}\GeV^2,\notag\\
(M_{K^0}^2)_\text{EM}&=7(2)(17)\times 10^{-6}\GeV^2\notag\\
&=7(17)\times 10^{-6}\GeV^2,\notag\\
(\Delta M_K^2)_\text{EM}&=2.12(2)(17)\times 10^{-3}\GeV^2\notag\\
&=2.12(18)\times 10^{-3}\GeV^2,\notag\\
(\Delta M_\pi^2)_\text{EM}&=1.3(3)\times10^{-3}\GeV^2,\notag\\
\eps&= 0.63(40). \label{eq:Cottingham-results}
\end{align}
The errors cover only the uncertainty in the elastic contributions, not the additional uncertainty from inelastic corrections. The second uncertainty for the kaons contains the variation of the UFD parameters as well as an uncertainty from the asymptotic continuation, all added linearly to account for possible correlations between the UFD and asymptotic uncertainties. The latter gives the dominant effect in this application, and has been estimated by varying the input for the pion VFF above $s=(2\GeV)^2$, changing the matching point to an asymptotic form of the isovector kaon form factor $F_K^v(-s) \asymp a/(b+s)\times (s/(b+s))^n$ for virtualities between $1\GeV^2$ and $10\GeV^2$, and varying the exponent $n=0,1$ to assess the impact of terms beyond the asymptotic $F_K(-s) \asymp 1/s$ behavior. Moreover, the final uncertainty from the asymptotic continuation has been inflated by a factor $2$ to account for a similar effect that could arise in the isoscalar contribution. For the pion the uncertainty is obtained from the error bands provided in Ref.~\cite{Colangelo:2018mtw} and also covers the uncertainty due to the asymptotic continuation.

In the end, the uncertainty in $\eps$ is dominated by the pion contribution. In part, this is due to the fact that the integration kernel gives a larger contribution from low virtualities for the kaon, resulting in a smaller relative uncertainty, but the more precise timelike data for the pion VFF also allow for a more detailed study of inelastic effects in the unitarity relation and thus a more robust error estimate for the continuation into the spacelike region.

\bsp
Equation~\eqref{eq:Cottingham-results} corresponds to the linear pion mass difference $\big(\mpc-\mpn\big)_{\text{EM}} = 4.8(1.1)\MeV$, which numerically saturates the experimental value $\mpc-\mpn = 4.59\MeV$ (the strong mass difference is estimated to give only a small contribution $\big(\mpc-\mpn\big)_{\text{str}} =0.17(3)\MeV$~\cite{Gasser:1984gg}).
\esp

Comparison to Eq.~\eqref{lattice} shows that this elastic estimate fully agrees with the lattice results, demonstrating that inelastic effects have to be smaller than the precision with which the elastic contributions can currently be evaluated. 

From Eq.~\eqref{eq:Cottingham-results}, we can extract the \textit{strong} kaon mass difference
\begin{equation}
    \big(\mKn^2-\mKc^2\big)_\text{str} = 6.02(18)\times10^{-3}\GeV^2,
\end{equation}
which is perfectly compatible with the result extracted from a dispersive analysis of $\eta\to3\pi$ decays, $\big(\mKn^2-\mKc^2\big)_\text{str} = 6.24(38)\times10^{-3}\GeV^2$~\cite{Colangelo:2018jxw}.
Similarly, if we convert our result for the strong kaon mass difference into a value for the quark mass ratio $Q$ according to Eq.~\eqref{eq:Q},\footnote{For the isospin symmetric masses $M_K^2$ and $M_\pi^2$ in Eq.~\eqref{eq:Q}, we use the average of charged and neutral squared kaon masses, subtracting the electromagnetic contribution as calculated here, and the neutral pion mass corrected for the strong mass shift~\cite{Gasser:1984gg}.} we find
\begin{equation}
    Q = 22.4(3), 
\end{equation}
again compatible with the value deduced from $\eta\to3\pi$, $Q=22.1(7)$~\cite{Colangelo:2018jxw}, as well as from several other analyses of the same decay~\cite{Kampf:2011wr,Guo:2016wsi,Colangelo:2016jmc,Albaladejo:2017hhj,Gan:2020aco}.  We wish to emphasize again that our errors here merely reflect the ones in the kaon and pion form factors, but not the omission of inelastic intermediate states in the Cottingham formula. 
In view of the large uncertainties, however, we expect that the assigned uncertainties also cover the omitted inelastic contributions, as is indeed the case for the pion mass difference. 

%-----------------------------------------------------------------------------------------
\subsection{Kaon-box contribution to HLbL scattering}
\label{sec:HLbL}
%-----------------------------------------------------------------------------------------
\begin{figure}[t]
    \centering
    \hspace{1cm}
    \begin{subfigure}[t]{0.15\textwidth}
        \centering
        \includegraphics[width=\textwidth]{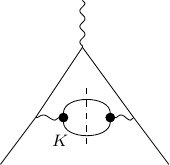}
    \end{subfigure}
    \hfill
    \begin{subfigure}[t]{0.15\textwidth}
        \centering
        \includegraphics[width=\textwidth]{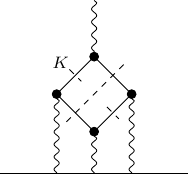}
    \end{subfigure}
    \hspace{1cm}
	\caption{Diagrammatic representation of the kaon contributions to HVP (left) and HLbL (right). The thick black dots denote kaon electromagnetic form factor insertions. Dashed lines indicate the $\bar K K$ cuts and, for HLbL scattering, the kaon pole in the $\gamma^*\gamma^{(*)}\to\bar K K$ subamplitudes.}
	\label{fig:Feynman-HLbL-HVP}
\end{figure}

The last two applications concern the anomalous magnetic moment of the muon, whose experimental world average~\cite{bennett:2006fi,Abi:2021gix,Albahri:2021ixb,Albahri:2021kmg,Albahri:2021mtf}
\begin{equation}
\label{exp}
a_\mu^\text{exp}=116\,592\,061(41)\times 10^{-11}
\end{equation}
currently displays a $4.2\sigma$ discrepancy to the Standard-Model (SM) prediction~\cite{Aoyama:2020ynm,Aoyama:2012wk,Aoyama:2019ryr,Czarnecki:2002nt,Gnendiger:2013pva,Davier:2017zfy,Keshavarzi:2018mgv,Colangelo:2018mtw,Hoferichter:2019mqg,Davier:2019can,Keshavarzi:2019abf,Hoid:2020xjs,Kurz:2014wya,Melnikov:2003xd,Colangelo:2014dfa,Colangelo:2014pva,Colangelo:2015ama,Masjuan:2017tvw,Colangelo:2017qdm,Colangelo:2017fiz,Hoferichter:2018dmo,Hoferichter:2018kwz,Gerardin:2019vio,Bijnens:2019ghy,Colangelo:2019lpu,Colangelo:2019uex,Blum:2019ugy,Colangelo:2014qya,Hoferichter:2021wyj}
\begin{equation}
\label{amuSM}
a_\mu^\text{SM}=116\,591\,810(43)\times 10^{-11}.
\end{equation}
Improvements are most pressing for the HVP contribution, for which the evaluation from $e^+e^-\to\text{hadrons}$ cross-section data that enters Eq.~\eqref{amuSM} stands in $2.1\sigma$ tension with the recent lattice-QCD calculation~\cite{Borsanyi:2020mff}, especially since the uncertainty in the leading-order HVP contribution~\cite{Aoyama:2020ynm,Davier:2017zfy,Keshavarzi:2018mgv,Colangelo:2018mtw,Hoferichter:2019mqg,Davier:2019can,Keshavarzi:2019abf,Hoid:2020xjs}
\begin{equation}
a_\mu^\text{HVP, LO}=6\,931(40)\times 10^{-11}
\end{equation}
emerges as the limiting factor in the total SM prediction. A large part of the uncertainty arises from systematic tensions among the $2\pi$ data sets, most notably between the BaBar~\cite{BaBar:2012bdw} and KLOE~\cite{KLOE-2:2017fda} data. It is therefore the $2\pi$ channel that receives most attention~\cite{Abbiendi:2022liz}, including when scrutinizing the consequences of the tension with Ref.~\cite{Borsanyi:2020mff}, see Refs.~\cite{Crivellin:2020zul,Keshavarzi:2020bfy,Malaescu:2020zuc,Colangelo:2020lcg}. However, compared to the final precision anticipated at the Fermilab experiment, $\Delta a_\mu^\text{exp}\text{[E989]}=16\times 10^{-11}$~\cite{Muong-2:2015xgu}, also the subleading channels are sizable, among them $e^+e^-\to K^+K^-$ and $e^+e^-\to K_SK_L$; cf.\ Fig.~\ref{fig:Feynman-HLbL-HVP}~(left). We will study their contributions in the vicinity of the $\phi$ region in Sec.~\ref{sec:HVP}. 

First, we turn to the HLbL contribution~\cite{Aoyama:2020ynm,Melnikov:2003xd,Colangelo:2014dfa,Colangelo:2014pva,Colangelo:2015ama,Masjuan:2017tvw,Colangelo:2017qdm,Colangelo:2017fiz,Hoferichter:2018dmo,Hoferichter:2018kwz,Gerardin:2019vio,Bijnens:2019ghy,Colangelo:2019lpu,Colangelo:2019uex,Pauk:2014rta,Danilkin:2016hnh,Jegerlehner:2017gek,Knecht:2018sci,Eichmann:2019bqf,Roig:2019reh,Blum:2019ugy}
\begin{equation}
    \label{eq:HLbL}
    a_\mu^\text{HLbL}=90(17)\times 10^{-11},
\end{equation}
in which case the uncertainty is dominated by subleading contributions beyond pseudoscalar poles and two-meson cuts, see Refs.~\cite{Holz:2015tcg,Hoferichter:2020lap,Ludtke:2020moa,Bijnens:2020xnl,Bijnens:2021jqo,Zanke:2021wiq,Chao:2021tvp,Danilkin:2021icn,Colangelo:2021nkr,Holz:2022hwz} for recent work in this direction. Meanwhile, $\bar K K$ states prove to be appreciably suppressed compared to their $\pi\pi$ analog, which is true for both the leading box contributions (estimated in VMD~\cite{Aoyama:2020ynm} or DS equations~\cite{Eichmann:2019bqf,Miramontes:2021exi}) and rescattering corrections~\cite{Danilkin:2021icn}, in the latter case despite the strong coupling between the $\pi\pi$ and $\bar K K$ channels via the $f_0(980)$. While the overall effect is thus known to be small, we can confirm these conclusions by means of our data-driven evaluation of the kaon form factors.

The pion-box contribution, defined dispersively in a double-spectral representation of HLbL scattering, was shown in Ref.~\cite{Colangelo:2015ama} to coincide with the scalar-QED one-loop expression, multiplied by pion electromagnetic form factors for the three virtual photons. The same applies to kaon intermediate states, see Fig.~\ref{fig:Feynman-HLbL-HVP}~(right): in the master formula for the contribution to $a_\mu$~\cite{Colangelo:2017fiz}, the HLbL scalar functions are of the form
\begin{align}
	\bar\Pi_i^{K\text{-box}}(Q_1^2,Q_2^2,Q_3^2) &= F_K(Q_1^2) F_K(Q_2^2) F_K(Q_3^2)\\
	&\times\frac{1}{16\pi^2} \int_0^1 \dd x \int_0^{1-x} \dd y \, I_i^K(x,y) ,\notag
\end{align}
where $Q_i^2$ denote the spacelike photon virtualities and the scalar-QED Feynman integrals are analogous to the pion case~\cite{Colangelo:2017fiz}. Thus, the charged- and neutral-kaon boxes can be obtained from the expressions for the pion box by simply replacing the meson mass and electromagnetic form factors. Diagrammatically, the box contributions correspond to two-meson cuts in which the $\gamma^*\gamma^{(*)}\to\pi\pi/\bar K K$ subamplitudes~\cite{GarciaMartin:2010cw,Hoferichter:2011wk,Moussallam:2013una,Danilkin:2018qfn,Hoferichter:2019nlq,Danilkin:2019opj} are further reduced to the respective poles. As in the case of the Cottingham formula, these contributions are fully determined by the electromagnetic form factors. 

\begin{table}[t]
\renewcommand{\arraystretch}{1.3}
	\begin{tabular}{lrr}
		\toprule
		 & $a_\mu^{K^\pm\text{-box}}\times 10^{11}$ & $a_\mu^{K^0\text{-box}}\times 10^{15}$\\\midrule
		VMD~\cite{Aoyama:2020ynm} & $-0.50$ & $-1.2$\\
		DS~\cite{Eichmann:2019bqf} & $-0.48(2)(4)$ & \\
		DS~\cite{Miramontes:2021exi} & $-0.48(4)$ & \\
		This work & $-0.484(5)(10)$ & $-0.5(2)(4)$\\
		\bottomrule
	\end{tabular}
	\centering
	\caption{Results for the charged and neutral kaon-box contributions to HLbL scattering.}
	\label{tab:kaon_box}
\renewcommand{\arraystretch}{1.0}
\end{table}

The numerical results for the kaon boxes with our data-driven form-factor input, shown in the spacelike region in Fig.~\ref{fig:F_K_full}, are in good agreement with previous estimates, see Table~\ref{tab:kaon_box}. The uncertainties have been obtained in the same way as in Sec.~\ref{sec:Dashen}. While in the case of the pion box about $95\%$ of the contribution is generated for photon virtualities below $1\GeV$,  due to the heavier kaon mass the charged kaon box is saturated to only $74\%$ by this energy region. For the neutral kaon box, this low-energy region is responsible for only about $25\%$ of the contribution due to the vanishing of the form factors at $Q_i^2=0$. The sensitivity to larger virtualities explains the large relative uncertainty. While the neutral-kaon box is numerically irrelevant for $(g-2)_\mu$, even the charged-kaon box,
\begin{equation}
\label{charged_box}
a_\mu^{K\text{-box}}=-0.48(1)\times 10^{-11},
\end{equation}
is of little importance in view of the overall uncertainty of the HLbL contribution~\eqref{eq:HLbL}.

%-----------------------------------------------------------------------------------------
\subsection{HVP around the $\phi$ resonance}
\label{sec:HVP}
%-----------------------------------------------------------------------------------------

The HVP contribution to $a_\mu$ derives from the master formula~\cite{Bouchiat:1961lbg,Brodsky:1967sr}
\begin{align}
\label{amu_HVP}
 a_\mu^\text{HVP, LO}&=\bigg(\frac{\alpha m_\mu}{3\pi}\bigg)^2\int_{\mpn^2}^\infty \dd s \frac{\hat K(s)}{s^2}R_\text{had}(s),\notag\\
 R_\text{had}(s)&=\frac{3s}{4\pi\alpha^2}\sigma(e^+e^-\to\text{hadrons}),
\end{align}
where $\hat K(s)$ is a known kernel function and the hadronic cross section is understood to be photon inclusive, see Fig.~\ref{fig:Feynman-HLbL-HVP} (left) for the contribution from the $\bar K K$ cut. For the neutral channel, the cross section thus follows directly from the form factor via Eq.~\eqref{Born}, while for the charged kaons the FSR correction needs to be added back. The numerical results are shown in Table~\ref{tab:kaon_HVP}, in comparison to the direct integration from Refs.~\cite{Keshavarzi:2018mgv,Keshavarzi:2019abf}. 

\begin{table}[t]
\renewcommand{\arraystretch}{1.3}
	\begin{tabular}{lrr}
		\toprule
		& $K^+K^-$ & $K_SK_L$\\\midrule
		Refs.~\cite{Keshavarzi:2018mgv,Keshavarzi:2019abf}  & $181.2(1.7)$ & $119.7(1.8)$\\
		charged/neutral  &$184.5(2.0)$ &  $118.3(1.5)$\\
		charged shift & $184.3(2.0)$ & \\
		combined & $184.5(2.0)$ & $118.3(1.5)$\\
		BaBar~\cite{BaBar:2013jqz} & $182.5(2.2)$ & \\
		CMD-2~\cite{CMD-2:2003gqi} & & $117.2(2.4)$ \\
		CMD-3~\cite{Kozyrev:2017agm,CMD-3:2016nhy} & $192.6(3.9)$ & $119.5(2.2)$\\
		\multirow{2}{*}{SND~\cite{Achasov:2000am}} & \multirow{2}{*}{$166.7(11.9)$} & $119.0(5.1)$ \\[-2pt]
		&& $121.1(5.0)$\\[2pt]
		\bottomrule
	\end{tabular}
	\centering
	\caption{Results for the charged- and neutral-kaon contributions to HVP up to $\sqrt{s}\leq 1.05\GeV$, in units of $10^{-11}$.}
	\label{tab:kaon_HVP}
\renewcommand{\arraystretch}{1.0}
\end{table}

For the neutral channel, the difference between the fit scenarios~\ref{chargedneutral} and~\ref{combined} is minimal, in both cases the integral comes out lower by less than $1\sigma$ in comparison to Refs.~\cite{Keshavarzi:2018mgv,Keshavarzi:2019abf}. This small difference mostly traces back to the use of a linear interpolation therein, given that the data points in the neutral channel are more clustered than in the charged one, see Fig.~\ref{fig:fits}. Accordingly, the same effect does not play a role in the charged channel~\cite{Keshavarzi2022}. In this case, we observe that the outcomes of the fit variants~\ref{chargedneutral} and~\ref{combined} are still well compatible, 
but larger by more than $1\sigma$ in comparison to Refs.~\cite{Keshavarzi:2018mgv,Keshavarzi:2019abf}.\footnote{Assuming uncorrelated errors, the difference would be $1.3\sigma$, but this is likely an underestimate since, due to the use of the same timelike data sets, the uncertainties are expected to be strongly correlated.}  

This difference is ultimately a manifestation of the tensions between the BaBar and CMD-3 data sets ($2.3\sigma$ for the HVP integral): in our approach, the cross section is constrained to follow the shape of the $\phi$ resonance in all fit variants, in such a way that the combined HVP integral comes out very close to the naive average of the individual data sets, despite the underlying tension, and, as shown in Table~\ref{tab:fit_parameters}, with a reasonable $\chi^2/\text{dof}$ even for the combined charged fit (albeit clearly worse than in the neutral channel, as long as no shift in energy calibration is allowed). Performing instead local averages of the data can obfuscate this global shape if inconsistencies are present among the data sets, and indeed a fit of our representation to the data combination of Refs.~\cite{Keshavarzi:2018mgv,Keshavarzi:2019abf} displays a $\chi^2/\text{dof}>2$.\footnote{We thank Alex Keshavarzi for making the combined cross-section data from Refs.~\cite{Keshavarzi:2018mgv,Keshavarzi:2019abf} available to us.} In particular, the increase in the $\chi^2/\text{dof}$ occurs because
the peak cross section is no longer compatible with the cross-section values in the tails of the resonance, to the effect that a linear interpolation gives the HVP integral as in Refs.~\cite{Keshavarzi:2018mgv,Keshavarzi:2019abf}, while enforcing our fit function does produce a larger value very close to our combined fit.  

Our final results are 
\begin{align}
    \label{HVP_result}
 a_\mu^\text{HVP}[K^+K^-, \leq 1.05\GeV]&=184.5(2.0)\times 10^{-11},\notag\\ a_\mu^\text{HVP}[K_SK_L, \leq 1.05\GeV]&=118.3(1.5)\times 10^{-11},   
\end{align}
quoted, as for the other applications before, from the combined fit~\ref{combined}. In this way, we implement the maximum amount of independent constraints available, emphasizing the complementarity of our results to the direct integration of the cross-section data~\cite{Davier:2017zfy,Keshavarzi:2018mgv,Davier:2019can,Keshavarzi:2019abf}. 
In particular, demanding universality of the $\phi$ pole parameters implies that part of the tension in the charged data base propagates into the result for the neutral channel via the scale factor of the fit, but the numerical impact is minimal, see Table~\ref{tab:kaon_HVP}.
Since, in this fit scenario, the numbers for the two channels become correlated via the $\phi$ pole parameters, we also quote the sum
\begin{align}
    \label{HVP_result_combined}
  &a_\mu^\text{HVP}[K^+K^-+K_SK_L, \leq 1.05\GeV]\notag\\
  &=302.8(2.5) \times 10^{-11}.
\end{align}

%-----------------------------------------------------------------------------------------
\section{Summary and conclusions}
\label{sec:summary}
%-----------------------------------------------------------------------------------------

In this paper we presented a comprehensive analysis of the electromagnetic form factors of the kaon, including all available constraints from dispersion relations and both time- and spacelike data. This approach has the advantage that a host of experimental constraints can be considered simultaneously, including (i) input from the $P$-wave of $\pi\pi\to\bar K K$ and the pion electromagnetic form factor, which together yield the bulk of the spectral function of the isovector kaon form factor, with a result that can be validated by data for $\tau^-\to K^- K_S\nu_\tau$, (ii) timelike data for $e^+e^-\to\bar K K$, which determine the residue of the $\phi$ resonance in the isoscalar part of the form factor, and (iii) spacelike data for charged-kaon--electron scattering, which help constrain the residue of the $\omega$, as not directly accessible in the timelike region. Further (indirect) constraints on the charge radius of the neutral kaon could be imposed, but for the reasons given in Sec.~\ref{sec:charge} we opt to provide an independent determination instead. Dispersion relations are most useful for the isovector form factor, leading to a model-independent implementation of the $\rho$ meson in terms of $2\pi$ intermediate states, while for the isoscalar spectral function, dominated by $3\pi$ and $\bar K K$ contributions, a parameterization in terms of the narrow $\omega$ and $\phi$ resonances proves sufficient. We considered several fit variants for the timelike data, described in~\ref{individual}--\ref{combined} in Sec.~\ref{sec:fits}, to account for tensions in the charged-channel data base. In particular, we studied to which extent the subsequent applications are affected.    

The results obtained along these lines for the kaon electromagnetic form factors enter in a number of applications. Besides the $\phi$ resonance parameters---the global fit leads to the values given in Eq.~\eqref{phi_parameters_our_fit}---these are: 
\begin{enumerate}
    \item Charge radii for charged and neutral kaon, see Sec.~\ref{sec:charge} and Eq.~\eqref{charge_radii_main_result} for the main result: for the charged kaon, our result lies $1.3\sigma$ above the current PDG average, but is considerably more precise, thanks to the inclusion of timelike data and dispersive constraints on the isovector form factor. The same is true for the neutral kaon, in which case our result lies $1.6\sigma$ above the PDG average (dominated by $K_L\to\pi^+\pi^- e^+e^-$), closer to the central value from $K^0$--electron scattering.
    \item Corrections to Dashen's theorem, see Sec.~\ref{sec:Dashen} and Eq.~\eqref{eq:Cottingham-results} for the main result: the spacelike kaon form factor determines the elastic contribution to the electromagnetic mass shift via the Cottingham formula, and thus the comparison to the analog formula for the pion entails a prediction for the corrections to Dashen's theorem. Our result is in perfect agreement with lattice QCD and extractions from $\eta\to3\pi$, demonstrating that the inelastic effects in the evaluation of the Cottingham formula are smaller than the current uncertainties in the elastic contribution. 
    \item Kaon-box contributions to HLbL scattering, see Sec.~\ref{sec:HLbL} and Eq.~\eqref{charged_box} for the main result: the spacelike form factor also arises in the evaluation of two-kaon intermediate states to HLbL scattering, so-called kaon-box contributions. Our result agrees with previous calculations, but provides a data-driven estimate of the uncertainty. 
    \item Two-kaon contributions to HVP, see Sec.~\ref{sec:HVP} and Eq.~\eqref{HVP_result} for the main result: the timelike data around the $\phi$ resonance, used to constrain the $\phi$ parameters in the isoscalar form factor, dominate the contribution to the HVP integral in the same energy region. Our evaluation delineates the potential impact of dispersive constraints and input from other kinematic regions, and also allows for a more detailed study of the consequences of the tensions in the charged-channel data base. In general, our results are in good agreement with previous evaluations using a direct integration of the data, but we find that, in the charged channel, the HVP integral increases by more than $1\sigma$ if the global shape of the $\phi$ resonance is enforced as it is in our dispersive representation. This difference ultimately reflects a tension between the BaBar and CMD-3 data for the $e^+e^-\to K^+ K^-$ channel. 
\end{enumerate}

%-----------------------------------------------------------------------------------------
\begin{acknowledgements}
\bsp
We thank Alex Keshavarzi for valuable communication, in particular for providing the comparison numbers from Refs.~\cite{Keshavarzi:2018mgv,Keshavarzi:2019abf},
and J\"urg Gasser and Christoph Hanhart for useful discussions.
Financial support by the SNSF (Project Nos.\ PCEFP2\_181117 and PCEFP2\_194272) and
the DFG through the funds provided to the Sino--German Collaborative Research Center TRR110 ``Symmetries and the Emergence of Structure in QCD'' (DFG Project-ID 196253076 -- TRR 110)
is gratefully acknowledged.
\esp
\end{acknowledgements}
%-----------------------------------------------------------------------------------------

\bibliographystyle{utphysmod}
\bibliography{Literature}

\providecommand{\href}[2]{#2}\begingroup\raggedright\begin{thebibliography}{100}

\bibitem{Omnes:1958hv}
R.~Omn{\`e}s, \href{http://dx.doi.org/10.1007/BF02747746}{Nuovo Cim. {\bfseries
  8}, 316 (1958)}.

\bibitem{DeTroconiz:2001rip}
J.~F. de~Troc\'oniz and F.~J. Yndur\'ain,
  \href{http://dx.doi.org/10.1103/PhysRevD.65.093001}{Phys. Rev. D {\bfseries
  65}, 093001 (2002)}
  [\href{https://arxiv.org/abs/hep-ph/0106025}{{arXiv:hep-ph/0106025}}].

\bibitem{Leutwyler:2002hm}
H.~Leutwyler, 2002.
\newblock \href{https://arxiv.org/abs/hep-ph/0212324}{{arXiv:hep-ph/0212324}}.

\bibitem{Colangelo:2003yw}
G.~Colangelo, \href{http://dx.doi.org/10.1016/j.nuclphysbps.2004.02.025}{Nucl.
  Phys. B Proc. Suppl. {\bfseries 131}, 185 (2004)}
  [\href{https://arxiv.org/abs/hep-ph/0312017}{{arXiv:hep-ph/0312017}}].

\bibitem{deTroconiz:2004yzs}
J.~F. de~Troc\'oniz and F.~J. Yndur\'ain,
  \href{http://dx.doi.org/10.1103/PhysRevD.71.073008}{Phys. Rev. D {\bfseries
  71}, 073008 (2005)}
  [\href{https://arxiv.org/abs/hep-ph/0402285}{{arXiv:hep-ph/0402285}}].

\bibitem{Ananthanarayan:2013zua}
B.~Ananthanarayan, I.~Caprini, D.~Das, and I.~Sentitemsu~Imsong,
  \href{http://dx.doi.org/10.1103/PhysRevD.89.036007}{Phys. Rev. D {\bfseries
  89}, 036007 (2014)} [\href{https://arxiv.org/abs/1312.5849}{{arXiv:1312.5849
  [hep-ph]}}].

\bibitem{Ananthanarayan:2016mns}
B.~Ananthanarayan, I.~Caprini, D.~Das, and I.~Sentitemsu~Imsong,
  \href{http://dx.doi.org/10.1103/PhysRevD.93.116007}{Phys. Rev. D {\bfseries
  93}, 116007 (2016)}
  [\href{https://arxiv.org/abs/1605.00202}{{arXiv:1605.00202 [hep-ph]}}].

\bibitem{Hoferichter:2016duk}
M.~Hoferichter, B.~Kubis, J.~Ruiz~de Elvira, H.-W. Hammer, and U.-G.
  Mei\ss{}ner, \href{http://dx.doi.org/10.1140/epja/i2016-16331-7}{Eur. Phys.
  J. A {\bfseries 52}, 331 (2016)}
  [\href{https://arxiv.org/abs/1609.06722}{{arXiv:1609.06722 [hep-ph]}}].

\bibitem{Hanhart:2016pcd}
C.~Hanhart, S.~Holz, B.~Kubis, A.~Kup\'s\'c, A.~Wirzba, and C.-W. Xiao,
  \href{http://dx.doi.org/10.1140/epjc/s10052-017-4651-x}{Eur. Phys. J. C
  {\bfseries 77}, 98 (2017)}
  [\href{https://arxiv.org/abs/1611.09359}{{arXiv:1611.09359 [hep-ph]}}],
  [Erratum: Eur. Phys. J. C {\bf 78}, 450 (2018)].

\bibitem{Colangelo:2018mtw}
G.~Colangelo, M.~Hoferichter, and P.~Stoffer,
  \href{http://dx.doi.org/10.1007/JHEP02(2019)006}{JHEP {\bfseries 02}, 006
  (2019)} [\href{https://arxiv.org/abs/1810.00007}{{arXiv:1810.00007
  [hep-ph]}}].

\bibitem{Ananthanarayan:2018nyx}
B.~Ananthanarayan, I.~Caprini, and D.~Das,
  \href{http://dx.doi.org/10.1103/PhysRevD.98.114015}{Phys. Rev. D {\bfseries
  98}, 114015 (2018)}
  [\href{https://arxiv.org/abs/1810.09265}{{arXiv:1810.09265 [hep-ph]}}].

\bibitem{Davier:2019can}
M.~Davier, A.~Hoecker, B.~Malaescu, and Z.~Zhang,
  \href{http://dx.doi.org/10.1140/epjc/s10052-020-7792-2}{Eur. Phys. J. C
  {\bfseries 80}, 241 (2020)}
  [\href{https://arxiv.org/abs/1908.00921}{{arXiv:1908.00921 [hep-ph]}}],
  [Erratum: Eur. Phys. J. C {\bf 80}, 410 (2020)].

\bibitem{Blatnik:1978wj}
S.~Blatnik, J.~Stahov, and C.~B. Lang,
  \href{http://dx.doi.org/10.1007/BF02725742}{Lett. Nuovo Cim. {\bfseries 24},
  39 (1979)}.

\bibitem{BaBar:2018qry}
J.~P. Lees {\em et~al.} [BaBar Collaboration],
  \href{http://dx.doi.org/10.1103/PhysRevD.98.032010}{Phys. Rev. D {\bfseries
  98}, 032010 (2018)}
  [\href{https://arxiv.org/abs/1806.10280}{{arXiv:1806.10280 [hep-ex]}}].

\bibitem{Achasov:2000am}
M.~N. Achasov {\em et~al.},
  \href{http://dx.doi.org/10.1103/PhysRevD.63.072002}{Phys. Rev. D {\bfseries
  63}, 072002 (2001)}
  [\href{https://arxiv.org/abs/hep-ex/0009036}{{arXiv:hep-ex/0009036}}].

\bibitem{CMD-2:2008fsu}
R.~R. Akhmetshin {\em et~al.} [CMD-2 Collaboration],
  \href{http://dx.doi.org/10.1016/j.physletb.2008.09.053}{Phys. Lett. B
  {\bfseries 669}, 217 (2008)}
  [\href{https://arxiv.org/abs/0804.0178}{{arXiv:0804.0178 [hep-ex]}}].

\bibitem{BaBar:2013jqz}
J.~P. Lees {\em et~al.} [BaBar Collaboration],
  \href{http://dx.doi.org/10.1103/PhysRevD.88.032013}{Phys. Rev. D {\bfseries
  88}, 032013 (2013)} [\href{https://arxiv.org/abs/1306.3600}{{arXiv:1306.3600
  [hep-ex]}}].

\bibitem{Kozyrev:2017agm}
E.~A. Kozyrev {\em et~al.},
  \href{http://dx.doi.org/10.1016/j.physletb.2018.01.079}{Phys. Lett. B
  {\bfseries 779}, 64 (2018)}
  [\href{https://arxiv.org/abs/1710.02989}{{arXiv:1710.02989 [hep-ex]}}].

\bibitem{CMD-2:2003gqi}
R.~R. Akhmetshin {\em et~al.} [CMD-2 Collaboration],
  \href{http://dx.doi.org/10.1016/j.physletb.2003.10.108}{Phys. Lett. B
  {\bfseries 578}, 285 (2004)}
  [\href{https://arxiv.org/abs/hep-ex/0308008}{{arXiv:hep-ex/0308008}}].

\bibitem{CMD-3:2016nhy}
E.~A. Kozyrev {\em et~al.} [CMD-3 Collaboration],
  \href{http://dx.doi.org/10.1016/j.physletb.2016.07.003}{Phys. Lett. B
  {\bfseries 760}, 314 (2016)}
  [\href{https://arxiv.org/abs/1604.02981}{{arXiv:1604.02981 [hep-ex]}}].

\bibitem{Dally:1980dj}
E.~B. Dally {\em et~al.},
  \href{http://dx.doi.org/10.1103/PhysRevLett.45.232}{Phys. Rev. Lett.
  {\bfseries 45}, 232 (1980)}.

\bibitem{Amendolia:1986ui}
S.~R. Amendolia {\em et~al.},
  \href{http://dx.doi.org/10.1016/0370-2693(86)91407-3}{Phys. Lett. B
  {\bfseries 178}, 435 (1986)}.

\bibitem{BaBar:2014uwz}
J.~P. Lees {\em et~al.} [BaBar Collaboration],
  \href{http://dx.doi.org/10.1103/PhysRevD.89.092002}{Phys. Rev. D {\bfseries
  89}, 092002 (2014)} [\href{https://arxiv.org/abs/1403.7593}{{arXiv:1403.7593
  [hep-ex]}}].

\bibitem{Hoferichter:2019mqg}
M.~Hoferichter, B.-L. Hoid, and B.~Kubis,
  \href{http://dx.doi.org/10.1007/JHEP08(2019)137}{JHEP {\bfseries 08}, 137
  (2019)} [\href{https://arxiv.org/abs/1907.01556}{{arXiv:1907.01556
  [hep-ph]}}].

\bibitem{Hoid:2020xjs}
B.-L. Hoid, M.~Hoferichter, and B.~Kubis,
  \href{http://dx.doi.org/10.1140/epjc/s10052-020-08550-2}{Eur. Phys. J. C
  {\bfseries 80}, 988 (2020)}
  [\href{https://arxiv.org/abs/2007.12696}{{arXiv:2007.12696 [hep-ph]}}].

\bibitem{Zyla:2020zbs}
P.~A. Zyla {\em et~al.} [Particle Data Group],
  \href{http://dx.doi.org/10.1093/ptep/ptaa104}{PTEP {\bfseries 2020}, 083C01
  (2020)}.

\bibitem{NA48:2003pwz}
A.~Lai {\em et~al.} [NA48 Collaboration],
  \href{http://dx.doi.org/10.1140/epjc/s2003-01252-y}{Eur. Phys. J. C
  {\bfseries 30}, 33 (2003)}.

\bibitem{KTeV:2005eic}
E.~Abouzaid {\em et~al.} [KTeV Collaboration],
  \href{http://dx.doi.org/10.1103/PhysRevLett.96.101801}{Phys. Rev. Lett.
  {\bfseries 96}, 101801 (2006)}
  [\href{https://arxiv.org/abs/hep-ex/0508010}{{arXiv:hep-ex/0508010}}].

\bibitem{Foeth:1969it}
H.~Foeth {\em et~al.},
  \href{http://dx.doi.org/10.1016/0370-2693(69)90439-0}{Phys. Lett. B
  {\bfseries 30}, 276 (1969)}.

\bibitem{Dydak:1976bv}
F.~Dydak {\em et~al.},
  \href{http://dx.doi.org/10.1016/0550-3213(76)90098-5}{Nucl. Phys. B
  {\bfseries 102}, 253 (1976)}.

\bibitem{Molzon:1978py}
W.~R. Molzon, J.~Hoffnagle, J.~Roehrig, V.~L. Telegdi, B.~Winstein, S.~H.
  Aronson, G.~J. Bock, D.~Hedin, G.~B. Thomson, and A.~Gsponer,
  \href{http://dx.doi.org/10.1103/PhysRevLett.41.1213}{Phys. Rev. Lett.
  {\bfseries 41}, 1213 (1978)}, [Errata: Phys. Rev. Lett. {\bf 41}, 1523
  (1978); 1835 (1978)].

\bibitem{Cottingham:1963zz}
W.~N. Cottingham, \href{http://dx.doi.org/10.1016/0003-4916(63)90023-X}{Annals
  Phys. {\bfseries 25}, 424 (1963)}.

\bibitem{Dashen:1969eg}
R.~F. Dashen, \href{http://dx.doi.org/10.1103/PhysRev.183.1245}{Phys. Rev.
  {\bfseries 183}, 1245 (1969)}.

\bibitem{Aoyama:2020ynm}
T.~Aoyama {\em et~al.},
  \href{http://dx.doi.org/10.1016/j.physrep.2020.07.006}{Phys. Rept. {\bfseries
  887}, 1 (2020)} [\href{https://arxiv.org/abs/2006.04822}{{arXiv:2006.04822
  [hep-ph]}}].

\bibitem{Eichmann:2019bqf}
G.~Eichmann, C.~S. Fischer, and R.~Williams,
  \href{http://dx.doi.org/10.1103/PhysRevD.101.054015}{Phys. Rev. D {\bfseries
  101}, 054015 (2020)}
  [\href{https://arxiv.org/abs/1910.06795}{{arXiv:1910.06795 [hep-ph]}}].

\bibitem{Miramontes:2021exi}
A.~Miramontes, A.~Bashir, K.~Raya, and P.~Roig,
  \href{http://dx.doi.org/10.1103/PhysRevD.105.074013}{Phys. Rev. D {\bfseries
  105}, 074013 (2022)}
  [\href{https://arxiv.org/abs/2112.13916}{{arXiv:2112.13916 [hep-ph]}}].

\bibitem{Davier:2017zfy}
M.~Davier, A.~Hoecker, B.~Malaescu, and Z.~Zhang,
  \href{http://dx.doi.org/10.1140/epjc/s10052-017-5161-6}{Eur. Phys. J. C
  {\bfseries 77}, 827 (2017)}
  [\href{https://arxiv.org/abs/1706.09436}{{arXiv:1706.09436 [hep-ph]}}].

\bibitem{Keshavarzi:2018mgv}
A.~Keshavarzi, D.~Nomura, and T.~Teubner,
  \href{http://dx.doi.org/10.1103/PhysRevD.97.114025}{Phys. Rev. D {\bfseries
  97}, 114025 (2018)}
  [\href{https://arxiv.org/abs/1802.02995}{{arXiv:1802.02995 [hep-ph]}}].

\bibitem{Keshavarzi:2019abf}
A.~Keshavarzi, D.~Nomura, and T.~Teubner,
  \href{http://dx.doi.org/10.1103/PhysRevD.101.014029}{Phys. Rev. D {\bfseries
  101}, 014029 (2020)}
  [\href{https://arxiv.org/abs/1911.00367}{{arXiv:1911.00367 [hep-ph]}}].

\bibitem{Pelaez:2020gnd}
J.~R. Pel\'aez and A.~Rodas,
  \href{https://arxiv.org/abs/2010.11222}{{arXiv:2010.11222 [hep-ph]}}.

\bibitem{Buettiker:2003pp}
P.~B{\"u}ttiker, S.~Descotes-Genon, and B.~Moussallam,
  \href{http://dx.doi.org/10.1140/epjc/s2004-01591-1}{Eur. Phys. J. C
  {\bfseries 33}, 409 (2004)}
  [\href{https://arxiv.org/abs/hep-ph/0310283}{{arXiv:hep-ph/0310283}}].

\bibitem{Pelaez:2018qny}
J.~R. Pel\'aez and A.~Rodas,
  \href{http://dx.doi.org/10.1140/epjc/s10052-018-6296-9}{Eur. Phys. J. C
  {\bfseries 78}, 897 (2018)}
  [\href{https://arxiv.org/abs/1807.04543}{{arXiv:1807.04543 [hep-ph]}}].

\bibitem{BaBar:2009wpw}
B.~Aubert {\em et~al.} [BaBar Collaboration],
  \href{http://dx.doi.org/10.1103/PhysRevLett.103.231801}{Phys. Rev. Lett.
  {\bfseries 103}, 231801 (2009)}
  [\href{https://arxiv.org/abs/0908.3589}{{arXiv:0908.3589 [hep-ex]}}].

\bibitem{Breit:1936zzb}
G.~Breit and E.~Wigner, \href{http://dx.doi.org/10.1103/PhysRev.49.519}{Phys.
  Rev. {\bfseries 49}, 519 (1936)}.

\bibitem{Hoferichter:2015hva}
M.~Hoferichter, J.~Ruiz~de Elvira, B.~Kubis, and U.-G. Mei{\ss}ner,
  \href{http://dx.doi.org/10.1016/j.physrep.2016.02.002}{Phys. Rept. {\bfseries
  625}, 1 (2016)} [\href{https://arxiv.org/abs/1510.06039}{{arXiv:1510.06039
  [hep-ph]}}].

\bibitem{Hanhart:2012wi}
C.~Hanhart, \href{http://dx.doi.org/10.1016/j.physletb.2012.07.038}{Phys. Lett.
  B {\bfseries 715}, 170 (2012)}
  [\href{https://arxiv.org/abs/1203.6839}{{arXiv:1203.6839 [hep-ph]}}].

\bibitem{Zanke:2021wiq}
M.~Zanke, M.~Hoferichter, and B.~Kubis,
  \href{http://dx.doi.org/10.1007/JHEP07(2021)106}{JHEP {\bfseries 07}, 106
  (2021)} [\href{https://arxiv.org/abs/2103.09829}{{arXiv:2103.09829
  [hep-ph]}}].

\bibitem{Schneider:2012ez}
S.~P. Schneider, B.~Kubis, and F.~Niecknig,
  \href{http://dx.doi.org/10.1103/PhysRevD.86.054013}{Phys. Rev. D {\bfseries
  86}, 054013 (2012)} [\href{https://arxiv.org/abs/1206.3098}{{arXiv:1206.3098
  [hep-ph]}}].

\bibitem{Hoferichter:2012pm}
M.~Hoferichter, B.~Kubis, and D.~Sakkas,
  \href{http://dx.doi.org/10.1103/PhysRevD.86.116009}{Phys. Rev. D {\bfseries
  86}, 116009 (2012)} [\href{https://arxiv.org/abs/1210.6793}{{arXiv:1210.6793
  [hep-ph]}}].

\bibitem{Chernyak:1977as}
V.~L. Chernyak and A.~R. Zhitnitsky, JETP Lett. {\bfseries 25}, 510 (1977),
  [Zh. Eksp. Teor. Fiz. {\bf 25}, 544 (1977)].

\bibitem{Chernyak:1980dj}
V.~L. Chernyak and A.~R. Zhitnitsky, Sov. J. Nucl. Phys. {\bfseries 31}, 544
  (1980), [Yad. Fiz. {\bf 31}, 1053 (1980)].

\bibitem{Efremov:1978rn}
A.~V. Efremov and A.~V. Radyushkin,
  \href{http://dx.doi.org/10.1007/BF01032111}{Theor. Math. Phys. {\bfseries
  42}, 97 (1980)}, [Teor. Mat. Fiz. {\bf 42}, 147 (1980)].

\bibitem{Efremov:1979qk}
A.~V. Efremov and A.~V. Radyushkin,
  \href{http://dx.doi.org/10.1016/0370-2693(80)90869-2}{Phys. Lett. B
  {\bfseries 94}, 245 (1980)}.

\bibitem{Farrar:1979aw}
G.~R. Farrar and D.~R. Jackson,
  \href{http://dx.doi.org/10.1103/PhysRevLett.43.246}{Phys. Rev. Lett.
  {\bfseries 43}, 246 (1979)}.

\bibitem{Lepage:1979zb}
G.~P. Lepage and S.~J. Brodsky,
  \href{http://dx.doi.org/10.1016/0370-2693(79)90554-9}{Phys. Lett. B
  {\bfseries 87}, 359 (1979)}.

\bibitem{Lepage:1980fj}
G.~P. Lepage and S.~J. Brodsky,
  \href{http://dx.doi.org/10.1103/PhysRevD.22.2157}{Phys. Rev. D {\bfseries
  22}, 2157 (1980)}.

\bibitem{Gonzalez-Solis:2019iod}
S.~Gonz\`alez-Sol\'\i{}s and P.~Roig,
  \href{http://dx.doi.org/10.1140/epjc/s10052-019-6943-9}{Eur. Phys. J. C
  {\bfseries 79}, 436 (2019)}
  [\href{https://arxiv.org/abs/1902.02273}{{arXiv:1902.02273 [hep-ph]}}].

\bibitem{Hoferichter:2014vra}
M.~Hoferichter, B.~Kubis, S.~Leupold, F.~Niecknig, and S.~P. Schneider,
  \href{http://dx.doi.org/10.1140/epjc/s10052-014-3180-0}{Eur. Phys. J. C
  {\bfseries 74}, 3180 (2014)}
  [\href{https://arxiv.org/abs/1410.4691}{{arXiv:1410.4691 [hep-ph]}}].

\bibitem{Fang:2021wes}
S.-s. Fang, B.~Kubis, and A.~Kup\'s\'c,
  \href{http://dx.doi.org/10.1016/j.ppnp.2021.103884}{Prog. Part. Nucl. Phys.
  {\bfseries 120}, 103884 (2021)}
  [\href{https://arxiv.org/abs/2102.05922}{{arXiv:2102.05922 [hep-ph]}}].

\bibitem{Hoefer:2001mx}
A.~Hoefer, J.~Gluza, and F.~Jegerlehner,
  \href{http://dx.doi.org/10.1007/s100520200916}{Eur. Phys. J. C {\bfseries
  24}, 51 (2002)}
  [\href{https://arxiv.org/abs/hep-ph/0107154}{{arXiv:hep-ph/0107154}}].

\bibitem{Gluza:2002ui}
J.~Gluza, A.~Hoefer, S.~Jadach, and F.~Jegerlehner,
  \href{http://dx.doi.org/10.1140/epjc/s2003-01146-0}{Eur. Phys. J. C
  {\bfseries 28}, 261 (2003)}
  [\href{https://arxiv.org/abs/hep-ph/0212386}{{arXiv:hep-ph/0212386}}].

\bibitem{Czyz:2004rj}
H.~Czy\.z, A.~Grzeli{\'n}ska, J.~H. K{\"u}hn, and G.~Rodrigo,
  \href{http://dx.doi.org/10.1140/epjc/s2004-02103-1}{Eur. Phys. J. C
  {\bfseries 39}, 411 (2005)}
  [\href{https://arxiv.org/abs/hep-ph/0404078}{{arXiv:hep-ph/0404078}}].

\bibitem{Bystritskiy:2005ib}
Y.~M. Bystritskiy, E.~A. Kuraev, G.~V. Fedotovich, and F.~V. Ignatov,
  \href{http://dx.doi.org/10.1103/PhysRevD.72.114019}{Phys. Rev. D {\bfseries
  72}, 114019 (2005)}
  [\href{https://arxiv.org/abs/hep-ph/0505236}{{arXiv:hep-ph/0505236}}].

\bibitem{sommerfeld1921atombau}
A.~Sommerfeld, {\em Atombau und Spektrallinien}, F. Vieweg \& Sohn, 1921.

\bibitem{Gamow:1928zz}
G.~Gamow, \href{http://dx.doi.org/10.1007/BF01343196}{Z. Phys. {\bfseries 51},
  204 (1928)}.

\bibitem{Sakharov:1948plh}
A.~D. Sakharov, \href{http://dx.doi.org/10.1070/PU1991v034n05ABEH002492}{Zh.
  Eksp. Teor. Fiz. {\bfseries 18}, 631 (1948)}.

\bibitem{Novosibirsk}
S.~Eidelman, F.~Ignatov, and E.~Kozyrev, 2017.
\newblock Private communication, as quoted in Ref.~\cite{Keshavarzi:2018mgv}.

\bibitem{DAgostini:1993arp}
G.~D'Agostini, \href{http://dx.doi.org/10.1016/0168-9002(94)90719-6}{Nucl.
  Instrum. Meth. A {\bfseries 346}, 306 (1994)}.

\bibitem{SND}
S.~Eidelman and F.~Ignatov, 2017.
\newblock Private communication, as quoted in Ref.~\cite{Keshavarzi:2018mgv}.

\bibitem{Ball:2009qv}
R.~D. Ball, L.~Del~Debbio, S.~Forte, A.~Guffanti, J.~I. Latorre, J.~Rojo, and
  M.~Ubiali [NNPDF Collaboration],
  \href{http://dx.doi.org/10.1007/JHEP05(2010)075}{JHEP {\bfseries 05}, 075
  (2010)} [\href{https://arxiv.org/abs/0912.2276}{{arXiv:0912.2276 [hep-ph]}}].

\bibitem{Beloborodov:2019fmw}
K.~I. Beloborodov, V.~P. Druzhinin, and S.~I. Serednyakov,
  \href{http://dx.doi.org/10.1134/S1063776119080016}{J. Exp. Theor. Phys.
  {\bfseries 129}, 386 (2019)}.

\bibitem{Bramon:2000qe}
A.~Bramon, R.~Escribano, J.~L. Lucio~M., and G.~Pancheri,
  \href{http://dx.doi.org/10.1016/S0370-2693(00)00770-X}{Phys. Lett. B
  {\bfseries 486}, 406 (2000)}
  [\href{https://arxiv.org/abs/hep-ph/0003273}{{arXiv:hep-ph/0003273}}].

\bibitem{Flores-Baez:2008owd}
F.~V. Flores-Ba\'ez and G.~L\'opez~Castro,
  \href{http://dx.doi.org/10.1103/PhysRevD.78.077301}{Phys. Rev. D {\bfseries
  78}, 077301 (2008)} [\href{https://arxiv.org/abs/0810.4349}{{arXiv:0810.4349
  [hep-ph]}}].

\bibitem{Benayoun:2012etq}
M.~Benayoun, P.~David, L.~DelBuono, and F.~Jegerlehner,
  \href{http://dx.doi.org/10.1140/epjc/s10052-011-1848-2}{Eur. Phys. J. C
  {\bfseries 72}, 1848 (2012)}
  [\href{https://arxiv.org/abs/1106.1315}{{arXiv:1106.1315 [hep-ph]}}].

\bibitem{Klingl:1996by}
F.~Klingl, N.~Kaiser, and W.~Weise,
  \href{http://dx.doi.org/10.1007/s002180050167}{Z. Phys. A {\bfseries 356},
  193 (1996)}
  [\href{https://arxiv.org/abs/hep-ph/9607431}{{arXiv:hep-ph/9607431}}].

\bibitem{Hoferichter:2017ftn}
M.~Hoferichter, B.~Kubis, and M.~Zanke,
  \href{http://dx.doi.org/10.1103/PhysRevD.96.114016}{Phys. Rev. D {\bfseries
  96}, 114016 (2017)}
  [\href{https://arxiv.org/abs/1710.00824}{{arXiv:1710.00824 [hep-ph]}}].

\bibitem{Dax:2020dzg}
M.~Dax, D.~Stamen, and B.~Kubis,
  \href{http://dx.doi.org/10.1140/epjc/s10052-021-08951-x}{Eur. Phys. J. C
  {\bfseries 81}, 221 (2021)}
  [\href{https://arxiv.org/abs/2012.04655}{{arXiv:2012.04655 [hep-ph]}}].

\bibitem{Gasser:1984ux}
J.~Gasser and H.~Leutwyler,
  \href{http://dx.doi.org/10.1016/0550-3213(85)90493-6}{Nucl. Phys. B
  {\bfseries 250}, 517 (1985)}.

\bibitem{Bijnens:2002hp}
J.~Bijnens and P.~Talavera,
  \href{http://dx.doi.org/10.1088/1126-6708/2002/03/046}{JHEP {\bfseries 03},
  046 (2002)}
  [\href{https://arxiv.org/abs/hep-ph/0203049}{{arXiv:hep-ph/0203049}}].

\bibitem{Leutwyler:1996qg}
H.~Leutwyler, \href{http://dx.doi.org/10.1016/0370-2693(96)00386-3}{Phys. Lett.
  B {\bfseries 378}, 313 (1996)}
  [\href{https://arxiv.org/abs/hep-ph/9602366}{{arXiv:hep-ph/9602366}}].

\bibitem{Gasser:1984gg}
J.~Gasser and H.~Leutwyler,
  \href{http://dx.doi.org/10.1016/0550-3213(85)90492-4}{Nucl. Phys. B
  {\bfseries 250}, 465 (1985)}.

\bibitem{Bijnens:1993ae}
J.~Bijnens, \href{http://dx.doi.org/10.1016/0370-2693(93)90089-Z}{Phys. Lett. B
  {\bfseries 306}, 343 (1993)}
  [\href{https://arxiv.org/abs/hep-ph/9302217}{{arXiv:hep-ph/9302217}}].

\bibitem{Donoghue:1993hj}
J.~F. Donoghue, B.~R. Holstein, and D.~Wyler,
  \href{http://dx.doi.org/10.1103/PhysRevD.47.2089}{Phys. Rev. D {\bfseries
  47}, 2089 (1993)}.

\bibitem{Baur:1995ig}
R.~Baur and R.~Urech, \href{http://dx.doi.org/10.1103/PhysRevD.53.6552}{Phys.
  Rev. D {\bfseries 53}, 6552 (1996)}
  [\href{https://arxiv.org/abs/hep-ph/9508393}{{arXiv:hep-ph/9508393}}].

\bibitem{Ananthanarayan:2004qk}
B.~Ananthanarayan and B.~Moussallam,
  \href{http://dx.doi.org/10.1088/1126-6708/2004/06/047}{JHEP {\bfseries 06},
  047 (2004)}
  [\href{https://arxiv.org/abs/hep-ph/0405206}{{arXiv:hep-ph/0405206}}].

\bibitem{Aoki:2021kgd}
Y.~Aoki {\em et~al.}, \href{https://arxiv.org/abs/2111.09849}{{arXiv:2111.09849
  [hep-lat]}}.

\bibitem{Borsanyi:2014jba}
S.~Borsanyi {\em et~al.},
  \href{http://dx.doi.org/10.1126/science.1257050}{Science {\bfseries 347},
  1452 (2015)} [\href{https://arxiv.org/abs/1406.4088}{{arXiv:1406.4088
  [hep-lat]}}].

\bibitem{Giusti:2017dmp}
D.~Giusti, V.~Lubicz, C.~Tarantino, G.~Martinelli, F.~Sanfilippo, S.~Simula,
  and N.~Tantalo, \href{http://dx.doi.org/10.1103/PhysRevD.95.114504}{Phys.
  Rev. D {\bfseries 95}, 114504 (2017)}
  [\href{https://arxiv.org/abs/1704.06561}{{arXiv:1704.06561 [hep-lat]}}].

\bibitem{MILC:2018ddw}
S.~Basak {\em et~al.} [MILC Collaboration],
  \href{http://dx.doi.org/10.1103/PhysRevD.99.034503}{Phys. Rev. D {\bfseries
  99}, 034503 (2019)}
  [\href{https://arxiv.org/abs/1807.05556}{{arXiv:1807.05556 [hep-lat]}}].

\bibitem{Fodor:2016bgu}
Z.~Fodor, C.~Hoelbling, S.~Krieg, L.~Lellouch, T.~Lippert, A.~Portelli,
  A.~Sastre, K.~K. Szabo, and L.~Varnhorst,
  \href{http://dx.doi.org/10.1103/PhysRevLett.117.082001}{Phys. Rev. Lett.
  {\bfseries 117}, 082001 (2016)}
  [\href{https://arxiv.org/abs/1604.07112}{{arXiv:1604.07112 [hep-lat]}}].

\bibitem{Harari:1966mu}
H.~Harari, \href{http://dx.doi.org/10.1103/PhysRevLett.17.1303}{Phys. Rev.
  Lett. {\bfseries 17}, 1303 (1966)}.

\bibitem{Gasser:1974wd}
J.~Gasser and H.~Leutwyler,
  \href{http://dx.doi.org/10.1016/0550-3213(75)90493-9}{Nucl. Phys. B
  {\bfseries 94}, 269 (1975)}.

\bibitem{Gasser:1982ap}
J.~Gasser and H.~Leutwyler,
  \href{http://dx.doi.org/10.1016/0370-1573(82)90035-7}{Phys. Rept. {\bfseries
  87}, 77 (1982)}.

\bibitem{Walker-Loud:2012ift}
A.~Walker-Loud, C.~E. Carlson, and G.~A. Miller,
  \href{http://dx.doi.org/10.1103/PhysRevLett.108.232301}{Phys. Rev. Lett.
  {\bfseries 108}, 232301 (2012)}
  [\href{https://arxiv.org/abs/1203.0254}{{arXiv:1203.0254 [nucl-th]}}].

\bibitem{Thomas:2014dxa}
A.~Thomas, X.~Wang, and R.~Young,
  \href{http://dx.doi.org/10.1103/PhysRevC.91.015209}{Phys. Rev. C {\bfseries
  91}, 015209 (2015)} [\href{https://arxiv.org/abs/1406.4579}{{arXiv:1406.4579
  [nucl-th]}}].

\bibitem{Erben:2014hza}
F.~Erben, P.~Shanahan, A.~Thomas, and R.~Young,
  \href{http://dx.doi.org/10.1103/PhysRevC.90.065205}{Phys. Rev. C {\bfseries
  90}, 065205 (2014)} [\href{https://arxiv.org/abs/1408.6628}{{arXiv:1408.6628
  [nucl-th]}}].

\bibitem{Gasser:2015dwa}
J.~Gasser, M.~Hoferichter, H.~Leutwyler, and A.~Rusetsky,
  \href{http://dx.doi.org/10.1140/epjc/s10052-015-3580-9}{Eur. Phys. J. C
  {\bfseries 75}, 375 (2015)}
  [\href{https://arxiv.org/abs/1506.06747}{{arXiv:1506.06747 [hep-ph]}}],
  [Erratum: Eur. Phys. J. C {\bf 80}, 353 (2020)].

\bibitem{Gasser:2020mzy}
J.~Gasser, H.~Leutwyler, and A.~Rusetsky,
  \href{http://dx.doi.org/10.1016/j.physletb.2021.136087}{Phys. Lett. B
  {\bfseries 814}, 136087 (2021)}
  [\href{https://arxiv.org/abs/2003.13612}{{arXiv:2003.13612 [hep-ph]}}].

\bibitem{Gasser:2020hzn}
J.~Gasser, H.~Leutwyler, and A.~Rusetsky,
  \href{http://dx.doi.org/10.1140/epjc/s10052-020-08615-2}{Eur. Phys. J. C
  {\bfseries 80}, 1121 (2020)}
  [\href{https://arxiv.org/abs/2008.05806}{{arXiv:2008.05806 [hep-ph]}}].

\bibitem{Ecker:1988te}
G.~Ecker, J.~Gasser, A.~Pich, and E.~de~Rafael,
  \href{http://dx.doi.org/10.1016/0550-3213(89)90346-5}{Nucl. Phys. B
  {\bfseries 321}, 311 (1989)}.

\bibitem{Bardeen:1988zw}
W.~A. Bardeen, J.~Bijnens, and J.~M. G{\'e}rard,
  \href{http://dx.doi.org/10.1103/PhysRevLett.62.1343}{Phys. Rev. Lett.
  {\bfseries 62}, 1343 (1989)}.

\bibitem{Donoghue:1996zn}
J.~F. Donoghue and A.~F. P{\'e}rez,
  \href{http://dx.doi.org/10.1103/PhysRevD.55.7075}{Phys. Rev. D {\bfseries
  55}, 7075 (1997)}
  [\href{https://arxiv.org/abs/hep-ph/9611331}{{arXiv:hep-ph/9611331}}].

\bibitem{Cirigliano:2020dmx}
V.~Cirigliano, W.~Dekens, J.~de~Vries, M.~Hoferichter, and E.~Mereghetti,
  \href{http://dx.doi.org/10.1103/PhysRevLett.126.172002}{Phys. Rev. Lett.
  {\bfseries 126}, 172002 (2021)}
  [\href{https://arxiv.org/abs/2012.11602}{{arXiv:2012.11602 [nucl-th]}}].

\bibitem{Cirigliano:2021qko}
V.~Cirigliano, W.~Dekens, J.~de~Vries, M.~Hoferichter, and E.~Mereghetti,
  \href{http://dx.doi.org/10.1007/JHEP05(2021)289}{JHEP {\bfseries 05}, 289
  (2021)} [\href{https://arxiv.org/abs/2102.03371}{{arXiv:2102.03371
  [nucl-th]}}].

\bibitem{Colangelo:2015ama}
G.~Colangelo, M.~Hoferichter, M.~Procura, and P.~Stoffer,
  \href{http://dx.doi.org/10.1007/JHEP09(2015)074}{JHEP {\bfseries 09}, 074
  (2015)} [\href{https://arxiv.org/abs/1506.01386}{{arXiv:1506.01386
  [hep-ph]}}].

\bibitem{Colangelo:2018jxw}
G.~Colangelo, S.~Lanz, H.~Leutwyler, and E.~Passemar,
  \href{http://dx.doi.org/10.1140/epjc/s10052-018-6377-9}{Eur. Phys. J. C
  {\bfseries 78}, 947 (2018)}
  [\href{https://arxiv.org/abs/1807.11937}{{arXiv:1807.11937 [hep-ph]}}].

\bibitem{Kampf:2011wr}
K.~Kampf, M.~Knecht, J.~Novotn\'y, and M.~Zdr\'ahal,
  \href{http://dx.doi.org/10.1103/PhysRevD.84.114015}{Phys. Rev. D {\bfseries
  84}, 114015 (2011)} [\href{https://arxiv.org/abs/1103.0982}{{arXiv:1103.0982
  [hep-ph]}}].

\bibitem{Guo:2016wsi}
P.~Guo, I.~V. Danilkin, C.~Fern{\'a}ndez-Ram{\'i}rez, V.~Mathieu, and A.~P.
  Szczepaniak, \href{http://dx.doi.org/10.1016/j.physletb.2017.05.092}{Phys.
  Lett. B {\bfseries 771}, 497 (2017)}
[\href{https://arxiv.org/abs/1608.01447}{{arXiv:1608.01447 [hep-ph]}}].
%%CITATION = ARXIV:1608.01447;%%.

\bibitem{Colangelo:2016jmc}
G.~Colangelo, S.~Lanz, H.~Leutwyler, and E.~Passemar,
  \href{http://dx.doi.org/10.1103/PhysRevLett.118.022001}{Phys. Rev. Lett.
  {\bfseries 118}, 022001 (2017)}
  [\href{https://arxiv.org/abs/1610.03494}{{arXiv:1610.03494 [hep-ph]}}].

\bibitem{Albaladejo:2017hhj}
M.~Albaladejo and B.~Moussallam,
  \href{http://dx.doi.org/10.1140/epjc/s10052-017-5052-x}{Eur. Phys. J. C
  {\bfseries 77}, 508 (2017)}
  [\href{https://arxiv.org/abs/1702.04931}{{arXiv:1702.04931 [hep-ph]}}].

\bibitem{Gan:2020aco}
L.~Gan, B.~Kubis, E.~Passemar, and S.~Tulin,
  \href{http://dx.doi.org/10.1016/j.physrep.2021.11.001}{Phys. Rept. {\bfseries
  945}, 2191 (2022)} [\href{https://arxiv.org/abs/2007.00664}{{arXiv:2007.00664
  [hep-ph]}}].

\bibitem{bennett:2006fi}
G.~W. Bennett {\em et~al.} [Muon $g-2$ Collaboration],
  \href{http://dx.doi.org/10.1103/PhysRevD.73.072003}{Phys. Rev. D {\bfseries
  73}, 072003 (2006)}
  [\href{https://arxiv.org/abs/hep-ex/0602035}{{arXiv:hep-ex/0602035}}].

\bibitem{Abi:2021gix}
B.~Abi {\em et~al.} [Muon $g-2$ Collaboration],
  \href{http://dx.doi.org/10.1103/PhysRevLett.126.141801}{Phys. Rev. Lett.
  {\bfseries 126}, 141801 (2021)}
  [\href{https://arxiv.org/abs/2104.03281}{{arXiv:2104.03281 [hep-ex]}}].

\bibitem{Albahri:2021ixb}
T.~Albahri {\em et~al.} [Muon $g-2$ Collaboration],
  \href{http://dx.doi.org/10.1103/PhysRevD.103.072002}{Phys. Rev. D {\bfseries
  103}, 072002 (2021)}
  [\href{https://arxiv.org/abs/2104.03247}{{arXiv:2104.03247 [hep-ex]}}].

\bibitem{Albahri:2021kmg}
T.~Albahri {\em et~al.} [Muon $g-2$ Collaboration],
  \href{http://dx.doi.org/10.1103/PhysRevA.103.042208}{Phys. Rev. A {\bfseries
  103}, 042208 (2021)}
  [\href{https://arxiv.org/abs/2104.03201}{{arXiv:2104.03201 [hep-ex]}}].

\bibitem{Albahri:2021mtf}
T.~Albahri {\em et~al.} [Muon $g-2$ Collaboration],
  \href{http://dx.doi.org/10.1103/PhysRevAccelBeams.24.044002}{Phys. Rev.
  Accel. Beams {\bfseries 24}, 044002 (2021)}
  [\href{https://arxiv.org/abs/2104.03240}{{arXiv:2104.03240
  [physics.acc-ph]}}].

\bibitem{Aoyama:2012wk}
T.~Aoyama, M.~Hayakawa, T.~Kinoshita, and M.~Nio,
  \href{http://dx.doi.org/10.1103/PhysRevLett.109.111808}{Phys. Rev. Lett.
  {\bfseries 109}, 111808 (2012)}
  [\href{https://arxiv.org/abs/1205.5370}{{arXiv:1205.5370 [hep-ph]}}].

\bibitem{Aoyama:2019ryr}
T.~Aoyama, T.~Kinoshita, and M.~Nio,
  \href{http://dx.doi.org/10.3390/atoms7010028}{Atoms {\bfseries 7}, 28
  (2019)}.

\bibitem{Czarnecki:2002nt}
A.~Czarnecki, W.~J. Marciano, and A.~Vainshtein,
  \href{http://dx.doi.org/10.1103/PhysRevD.67.073006}{Phys. Rev. D {\bfseries
  67}, 073006 (2003)}
  [\href{https://arxiv.org/abs/hep-ph/0212229}{{arXiv:hep-ph/0212229}}],
  [Erratum: Phys. Rev. D {\bf 73}, 119901 (2006)].

\bibitem{Gnendiger:2013pva}
C.~Gnendiger, D.~St\"ockinger, and H.~St\"ockinger-Kim,
  \href{http://dx.doi.org/10.1103/PhysRevD.88.053005}{Phys. Rev. D {\bfseries
  88}, 053005 (2013)} [\href{https://arxiv.org/abs/1306.5546}{{arXiv:1306.5546
  [hep-ph]}}].

\bibitem{Kurz:2014wya}
A.~Kurz, T.~Liu, P.~Marquard, and M.~Steinhauser,
  \href{http://dx.doi.org/10.1016/j.physletb.2014.05.043}{Phys. Lett. B
  {\bfseries 734}, 144 (2014)}
  [\href{https://arxiv.org/abs/1403.6400}{{arXiv:1403.6400 [hep-ph]}}].

\bibitem{Melnikov:2003xd}
K.~Melnikov and A.~Vainshtein,
  \href{http://dx.doi.org/10.1103/PhysRevD.70.113006}{Phys. Rev. D {\bfseries
  70}, 113006 (2004)}
  [\href{https://arxiv.org/abs/hep-ph/0312226}{{arXiv:hep-ph/0312226}}].

\bibitem{Colangelo:2014dfa}
G.~Colangelo, M.~Hoferichter, M.~Procura, and P.~Stoffer,
  \href{http://dx.doi.org/10.1007/JHEP09(2014)091}{JHEP {\bfseries 09}, 091
  (2014)} [\href{https://arxiv.org/abs/1402.7081}{{arXiv:1402.7081 [hep-ph]}}].

\bibitem{Colangelo:2014pva}
G.~Colangelo, M.~Hoferichter, B.~Kubis, M.~Procura, and P.~Stoffer,
  \href{http://dx.doi.org/10.1016/j.physletb.2014.09.021}{Phys. Lett. B
  {\bfseries 738}, 6 (2014)}
  [\href{https://arxiv.org/abs/1408.2517}{{arXiv:1408.2517 [hep-ph]}}].

\bibitem{Masjuan:2017tvw}
P.~Masjuan and P.~S{\'a}nchez-Puertas,
  \href{http://dx.doi.org/10.1103/PhysRevD.95.054026}{Phys. Rev. D {\bfseries
  95}, 054026 (2017)}
  [\href{https://arxiv.org/abs/1701.05829}{{arXiv:1701.05829 [hep-ph]}}].

\bibitem{Colangelo:2017qdm}
G.~Colangelo, M.~Hoferichter, M.~Procura, and P.~Stoffer,
  \href{http://dx.doi.org/10.1103/PhysRevLett.118.232001}{Phys. Rev. Lett.
  {\bfseries 118}, 232001 (2017)}
  [\href{https://arxiv.org/abs/1701.06554}{{arXiv:1701.06554 [hep-ph]}}].

\bibitem{Colangelo:2017fiz}
G.~Colangelo, M.~Hoferichter, M.~Procura, and P.~Stoffer,
  \href{http://dx.doi.org/10.1007/JHEP04(2017)161}{JHEP {\bfseries 04}, 161
  (2017)} [\href{https://arxiv.org/abs/1702.07347}{{arXiv:1702.07347
  [hep-ph]}}].

\bibitem{Hoferichter:2018dmo}
M.~Hoferichter, B.-L. Hoid, B.~Kubis, S.~Leupold, and S.~P. Schneider,
  \href{http://dx.doi.org/10.1103/PhysRevLett.121.112002}{Phys. Rev. Lett.
  {\bfseries 121}, 112002 (2018)}
  [\href{https://arxiv.org/abs/1805.01471}{{arXiv:1805.01471 [hep-ph]}}].

\bibitem{Hoferichter:2018kwz}
M.~Hoferichter, B.-L. Hoid, B.~Kubis, S.~Leupold, and S.~P. Schneider,
  \href{http://dx.doi.org/10.1007/JHEP10(2018)141}{JHEP {\bfseries 10}, 141
  (2018)} [\href{https://arxiv.org/abs/1808.04823}{{arXiv:1808.04823
  [hep-ph]}}].

\bibitem{Gerardin:2019vio}
A.~G\'erardin, H.~B. Meyer, and A.~Nyffeler,
  \href{http://dx.doi.org/10.1103/PhysRevD.100.034520}{Phys. Rev. D {\bfseries
  100}, 034520 (2019)}
  [\href{https://arxiv.org/abs/1903.09471}{{arXiv:1903.09471 [hep-lat]}}].

\bibitem{Bijnens:2019ghy}
J.~Bijnens, N.~Hermansson-Truedsson, and A.~Rodr\'\i{}guez-S\'anchez,
  \href{http://dx.doi.org/10.1016/j.physletb.2019.134994}{Phys. Lett. B
  {\bfseries 798}, 134994 (2019)}
  [\href{https://arxiv.org/abs/1908.03331}{{arXiv:1908.03331 [hep-ph]}}].

\bibitem{Colangelo:2019lpu}
G.~Colangelo, F.~Hagelstein, M.~Hoferichter, L.~Laub, and P.~Stoffer,
  \href{http://dx.doi.org/10.1103/PhysRevD.101.051501}{Phys. Rev. D {\bfseries
  101}, 051501 (2020)}
  [\href{https://arxiv.org/abs/1910.11881}{{arXiv:1910.11881 [hep-ph]}}].

\bibitem{Colangelo:2019uex}
G.~Colangelo, F.~Hagelstein, M.~Hoferichter, L.~Laub, and P.~Stoffer,
  \href{http://dx.doi.org/10.1007/JHEP03(2020)101}{JHEP {\bfseries 03}, 101
  (2020)} [\href{https://arxiv.org/abs/1910.13432}{{arXiv:1910.13432
  [hep-ph]}}].

\bibitem{Blum:2019ugy}
T.~Blum, N.~Christ, M.~Hayakawa, T.~Izubuchi, L.~Jin, C.~Jung, and C.~Lehner,
  \href{http://dx.doi.org/10.1103/PhysRevLett.124.132002}{Phys. Rev. Lett.
  {\bfseries 124}, 132002 (2020)}
  [\href{https://arxiv.org/abs/1911.08123}{{arXiv:1911.08123 [hep-lat]}}].

\bibitem{Colangelo:2014qya}
G.~Colangelo, M.~Hoferichter, A.~Nyffeler, M.~Passera, and P.~Stoffer,
  \href{http://dx.doi.org/10.1016/j.physletb.2014.06.012}{Phys. Lett. B
  {\bfseries 735}, 90 (2014)}
  [\href{https://arxiv.org/abs/1403.7512}{{arXiv:1403.7512 [hep-ph]}}].

\bibitem{Hoferichter:2021wyj}
M.~Hoferichter and T.~Teubner,
  \href{http://dx.doi.org/10.1103/PhysRevLett.128.112002}{Phys. Rev. Lett.
  {\bfseries 128}, 112002 (2022)}
  [\href{https://arxiv.org/abs/2112.06929}{{arXiv:2112.06929 [hep-ph]}}].

\bibitem{Borsanyi:2020mff}
S.~Borsanyi {\em et~al.},
  \href{http://dx.doi.org/10.1038/s41586-021-03418-1}{Nature {\bfseries 593},
  51 (2021)} [\href{https://arxiv.org/abs/2002.12347}{{arXiv:2002.12347
  [hep-lat]}}].

\bibitem{BaBar:2012bdw}
J.~P. Lees {\em et~al.} [BaBar Collaboration],
  \href{http://dx.doi.org/10.1103/PhysRevD.86.032013}{Phys. Rev. D {\bfseries
  86}, 032013 (2012)} [\href{https://arxiv.org/abs/1205.2228}{{arXiv:1205.2228
  [hep-ex]}}].

\bibitem{KLOE-2:2017fda}
A.~Anastasi {\em et~al.} [KLOE-2 Collaboration],
  \href{http://dx.doi.org/10.1007/JHEP03(2018)173}{JHEP {\bfseries 03}, 173
  (2018)} [\href{https://arxiv.org/abs/1711.03085}{{arXiv:1711.03085
  [hep-ex]}}].

\bibitem{Abbiendi:2022liz}
G.~Abbiendi {\em et~al.},
  \href{https://arxiv.org/abs/2201.12102}{{arXiv:2201.12102 [hep-ph]}}.

\bibitem{Crivellin:2020zul}
A.~Crivellin, M.~Hoferichter, C.~A. Manzari, and M.~Montull,
  \href{http://dx.doi.org/10.1103/PhysRevLett.125.091801}{Phys. Rev. Lett.
  {\bfseries 125}, 091801 (2020)}
  [\href{https://arxiv.org/abs/2003.04886}{{arXiv:2003.04886 [hep-ph]}}].

\bibitem{Keshavarzi:2020bfy}
A.~Keshavarzi, W.~J. Marciano, M.~Passera, and A.~Sirlin,
  \href{http://dx.doi.org/10.1103/PhysRevD.102.033002}{Phys. Rev. D {\bfseries
  102}, 033002 (2020)}
  [\href{https://arxiv.org/abs/2006.12666}{{arXiv:2006.12666 [hep-ph]}}].

\bibitem{Malaescu:2020zuc}
B.~Malaescu and M.~Schott,
  \href{http://dx.doi.org/10.1140/epjc/s10052-021-08848-9}{Eur. Phys. J. C
  {\bfseries 81}, 46 (2021)}
  [\href{https://arxiv.org/abs/2008.08107}{{arXiv:2008.08107 [hep-ph]}}].

\bibitem{Colangelo:2020lcg}
G.~Colangelo, M.~Hoferichter, and P.~Stoffer,
  \href{http://dx.doi.org/10.1016/j.physletb.2021.136073}{Phys. Lett. B
  {\bfseries 814}, 136073 (2021)}
  [\href{https://arxiv.org/abs/2010.07943}{{arXiv:2010.07943 [hep-ph]}}].

\bibitem{Muong-2:2015xgu}
J.~Grange {\em et~al.} [Muon $g-2$ Collaboration]
  [\href{https://arxiv.org/abs/1501.06858}{{arXiv:1501.06858
  [physics.ins-det]}}].

\bibitem{Pauk:2014rta}
V.~Pauk and M.~Vanderhaeghen,
  \href{http://dx.doi.org/10.1140/epjc/s10052-014-3008-y}{Eur. Phys. J. C
  {\bfseries 74}, 3008 (2014)}
  [\href{https://arxiv.org/abs/1401.0832}{{arXiv:1401.0832 [hep-ph]}}].

\bibitem{Danilkin:2016hnh}
I.~Danilkin and M.~Vanderhaeghen,
  \href{http://dx.doi.org/10.1103/PhysRevD.95.014019}{Phys. Rev. D {\bfseries
  95}, 014019 (2017)}
  [\href{https://arxiv.org/abs/1611.04646}{{arXiv:1611.04646 [hep-ph]}}].

\bibitem{Jegerlehner:2017gek}
F.~Jegerlehner, {\em {The Anomalous Magnetic Moment of the Muon}}, Springer,
  2017.

\bibitem{Knecht:2018sci}
M.~Knecht, S.~Narison, A.~Rabemananjara, and D.~Rabetiarivony,
  \href{http://dx.doi.org/10.1016/j.physletb.2018.10.048}{Phys. Lett. B
  {\bfseries 787}, 111 (2018)}
  [\href{https://arxiv.org/abs/1808.03848}{{arXiv:1808.03848 [hep-ph]}}].

\bibitem{Roig:2019reh}
P.~Roig and P.~S{\'a}nchez-Puertas,
  \href{http://dx.doi.org/10.1103/PhysRevD.101.074019}{Phys. Rev. D {\bfseries
  101}, 074019 (2020)}
  [\href{https://arxiv.org/abs/1910.02881}{{arXiv:1910.02881 [hep-ph]}}].

\bibitem{Holz:2015tcg}
S.~Holz, J.~Plenter, C.-W. Xiao, T.~Dato, C.~Hanhart, B.~Kubis, U.-G.
  Mei{\ss}ner, and A.~Wirzba,
  \href{http://dx.doi.org/10.1140/epjc/s10052-021-09661-0}{Eur. Phys. J. C
  {\bfseries 81}, 1002 (2021)}
  [\href{https://arxiv.org/abs/1509.02194}{{arXiv:1509.02194 [hep-ph]}}].

\bibitem{Hoferichter:2020lap}
M.~Hoferichter and P.~Stoffer,
  \href{http://dx.doi.org/10.1007/JHEP05(2020)159}{JHEP {\bfseries 05}, 159
  (2020)} [\href{https://arxiv.org/abs/2004.06127}{{arXiv:2004.06127
  [hep-ph]}}].

\bibitem{Ludtke:2020moa}
J.~L\"udtke and M.~Procura,
  \href{http://dx.doi.org/10.1140/epjc/s10052-020-08611-6}{Eur. Phys. J. C
  {\bfseries 80}, 1108 (2020)}
  [\href{https://arxiv.org/abs/2006.00007}{{arXiv:2006.00007 [hep-ph]}}].

\bibitem{Bijnens:2020xnl}
J.~Bijnens, N.~Hermansson-Truedsson, L.~Laub, and A.~Rodr\'iguez-S\'anchez,
  \href{http://dx.doi.org/10.1007/JHEP10(2020)203}{JHEP {\bfseries 10}, 203
  (2020)} [\href{https://arxiv.org/abs/2008.13487}{{arXiv:2008.13487
  [hep-ph]}}].

\bibitem{Bijnens:2021jqo}
J.~Bijnens, N.~Hermansson-Truedsson, L.~Laub, and A.~Rodr\'iguez-S\'anchez,
  \href{http://dx.doi.org/10.1007/JHEP04(2021)240}{JHEP {\bfseries 04}, 240
  (2021)} [\href{https://arxiv.org/abs/2101.09169}{{arXiv:2101.09169
  [hep-ph]}}].

\bibitem{Chao:2021tvp}
E.-H. Chao, R.~J. Hudspith, A.~G\'erardin, J.~R. Green, H.~B. Meyer, and
  K.~Ottnad, \href{http://dx.doi.org/10.1140/epjc/s10052-021-09455-4}{Eur.
  Phys. J. C {\bfseries 81}, 651 (2021)}
  [\href{https://arxiv.org/abs/2104.02632}{{arXiv:2104.02632 [hep-lat]}}].

\bibitem{Danilkin:2021icn}
I.~Danilkin, M.~Hoferichter, and P.~Stoffer,
  \href{http://dx.doi.org/10.1016/j.physletb.2021.136502}{Phys. Lett. B
  {\bfseries 820}, 136502 (2021)}
  [\href{https://arxiv.org/abs/2105.01666}{{arXiv:2105.01666 [hep-ph]}}].

\bibitem{Colangelo:2021nkr}
G.~Colangelo, F.~Hagelstein, M.~Hoferichter, L.~Laub, and P.~Stoffer,
  \href{http://dx.doi.org/10.1140/epjc/s10052-021-09513-x}{Eur. Phys. J. C
  {\bfseries 81}, 702 (2021)}
  [\href{https://arxiv.org/abs/2106.13222}{{arXiv:2106.13222 [hep-ph]}}].

\bibitem{Holz:2022hwz}
S.~Holz, C.~Hanhart, M.~Hoferichter, and B.~Kubis,
  \href{http://dx.doi.org/10.1140/epjc/s10052-022-10247-7}{Eur. Phys. J. C
  {\bfseries 82}, 434 (2022)}
  [\href{https://arxiv.org/abs/2202.05846}{{arXiv:2202.05846 [hep-ph]}}].

\bibitem{GarciaMartin:2010cw}
R.~Garc\'ia-Mart\'in and B.~Moussallam,
  \href{http://dx.doi.org/10.1140/epjc/s10052-010-1471-7}{Eur. Phys. J. C
  {\bfseries 70}, 155 (2010)}
  [\href{https://arxiv.org/abs/1006.5373}{{arXiv:1006.5373 [hep-ph]}}].

\bibitem{Hoferichter:2011wk}
M.~Hoferichter, D.~R. Phillips, and C.~Schat,
  \href{http://dx.doi.org/10.1140/epjc/s10052-011-1743-x}{Eur. Phys. J. C
  {\bfseries 71}, 1743 (2011)}
  [\href{https://arxiv.org/abs/1106.4147}{{arXiv:1106.4147 [hep-ph]}}].

\bibitem{Moussallam:2013una}
B.~Moussallam, \href{http://dx.doi.org/10.1140/epjc/s10052-013-2539-y}{Eur.
  Phys. J. C {\bfseries 73}, 2539 (2013)}
  [\href{https://arxiv.org/abs/1305.3143}{{arXiv:1305.3143 [hep-ph]}}].

\bibitem{Danilkin:2018qfn}
I.~Danilkin and M.~Vanderhaeghen,
  \href{http://dx.doi.org/10.1016/j.physletb.2018.12.047}{Phys. Lett. B
  {\bfseries 789}, 366 (2019)}
  [\href{https://arxiv.org/abs/1810.03669}{{arXiv:1810.03669 [hep-ph]}}].

\bibitem{Hoferichter:2019nlq}
M.~Hoferichter and P.~Stoffer,
  \href{http://dx.doi.org/10.1007/JHEP07(2019)073}{JHEP {\bfseries 07}, 073
  (2019)} [\href{https://arxiv.org/abs/1905.13198}{{arXiv:1905.13198
  [hep-ph]}}].

\bibitem{Danilkin:2019opj}
I.~Danilkin, O.~Deineka, and M.~Vanderhaeghen,
  \href{http://dx.doi.org/10.1103/PhysRevD.101.054008}{Phys. Rev. D {\bfseries
  101}, 054008 (2020)}
  [\href{https://arxiv.org/abs/1909.04158}{{arXiv:1909.04158 [hep-ph]}}].

\bibitem{Bouchiat:1961lbg}
C.~Bouchiat and L.~Michel,
  \href{http://dx.doi.org/10.1051/jphysrad:01961002202012101}{J. Phys. Radium
  {\bfseries 22}, 121 (1961)}.

\bibitem{Brodsky:1967sr}
S.~J. Brodsky and E.~de~Rafael,
  \href{http://dx.doi.org/10.1103/PhysRev.168.1620}{Phys. Rev. {\bfseries 168},
  1620 (1968)}.

\bibitem{Keshavarzi2022}
A.~Keshavarzi, 2022.
\newblock Private communication.

\end{thebibliography}\endgroup

\end{document}